\documentclass[11pt]{article}

\usepackage{acl}
\usepackage{times}
\usepackage{latexsym}

\usepackage[T1]{fontenc}
\usepackage[utf8]{inputenc}
\usepackage{microtype}
\usepackage{inconsolata}
\usepackage{graphicx}
\usepackage{subcaption} 
\usepackage{float}
\usepackage{amsmath,amssymb}
\newtheorem{lemma}{Lemma}
\newtheorem{theorem}{Theorem}
\usepackage{booktabs}
\usepackage{enumitem}
\usepackage{arydshln}
\usepackage{listings}
\usepackage{multirow}
\usepackage{algorithm}
\usepackage{algorithmicx}
\usepackage{algpseudocode}
\newcommand{\nosection}[1]{\vspace{2pt}\noindent\textbf{#1.}}
\usepackage{booktabs}
\usepackage{multirow}
\usepackage{xcolor}
\usepackage[table]{xcolor}
\usepackage{colortbl}
\usepackage{adjustbox}
\usepackage{hhline}

\newcommand{\result}[2]{
#1{\tiny\,$\pm$\,#2}%
}

\newcommand{\best}[2]{%
  \textbf{#1{\tiny\,$\pm$\,#2}}%
}

\newcommand{\second}[2]{%
  \underline{#1{\tiny\,$\pm$\,#2}}%
}
\newcommand{\ourscell}[1]{\cellcolor{cyan!9}#1}
\title{PGPO: Prototype-Guided Progressive Obfuscation for Privacy-Preserving LLM-Enhanced Recommendation}
\author{
  Shengxiang Lin \\
  Zhejiang University \\
  Hangzhou, China \\
  \texttt{linshengxiang@zju.edu.cn}
  \And
  Jiajie Su \\
  Zhejiang University \\
  Hangzhou, China \\
  \texttt{sujiajie@zju.edu.cn}
  \And
  Pengyang Zhou \\
  Zhejiang University \\
  Hangzhou, China \\
  \texttt{zhoupy@zju.edu.cn}
  \AND
  Xiang Chen \\
  Zhejiang University \\
  Hangzhou, China \\
  \texttt{chenxiang.zjucs@zju.edu.cn}
  \And
  Xiaolin Zheng$^{*}$ \\
  Zhejiang University \\
  Hangzhou, China \\
  \texttt{xlzheng@zju.edu.cn}
  \And
  Feng Tian$^{*}$ \\
  Xi'an Jiaotong University \\
  Xi'an, China \\
  \texttt{fengtian@mail.xjtu.edu.cn}
  \AND
  Chaochao Chen \\
  Zhejiang University \\
  Hangzhou, China \\
  \texttt{zjuccc@zju.edu.cn}
  \\
}

\begin{document}
\maketitle

\begingroup
\renewcommand{\thefootnote}{*}
\footnotetext{Corresponding Authors.}
\endgroup

\begin{abstract}
LLM-enhanced recommender systems encode item side information into semantic embeddings, but these embeddings may leak sensitive textual attributes through inversion attacks. Existing input-side defenses rely on trusted collection or local token perturbation, often disrupting the inter-item semantic topology needed for recommendation. To address it, we introduce \textbf{relational fidelity} as the key objective, requiring semantic shift for privacy while preserving relative item relations for utility. Based on this principle, we propose \textbf{prototype-guided progressive obfuscation (PGPO)}, which learns topology-aware obfuscation anchors for high-influence prototype words and propagates them through a static semantic graph. PGPO is optimized with GRPO using individual dissimilarity, structural consistency, and semantic constraint rewards. Experiments on two benchmarks show strong inversion defense with near-upper-bound recommendation performance. The code is available at \href{https://github.com/Shengxiang-Lin/PGPO}{https://github.com/Shengxiang-Lin/PGPO}.
\end{abstract}
\section{Introduction}
\label{sec:intro}

In recent years, large language models (LLMs) have leveraged their rich semantic understanding to provide high-quality embeddings for item side information~\citep{wu2024survey}. This approach effectively alleviates data sparsity and cold-start issues in recommendation. However, these embeddings may leak sensitive user information~\citep{song2020information,wang2025privacy}. Embedding inversion attacks can reconstruct the original text from public embeddings with high accuracy, and the risk increases when data is sent to uncontrolled third-party LLM APIs~\citep{tsai2026concept}.

\begin{figure}[t]
    \centering
    \includegraphics[width=\linewidth]{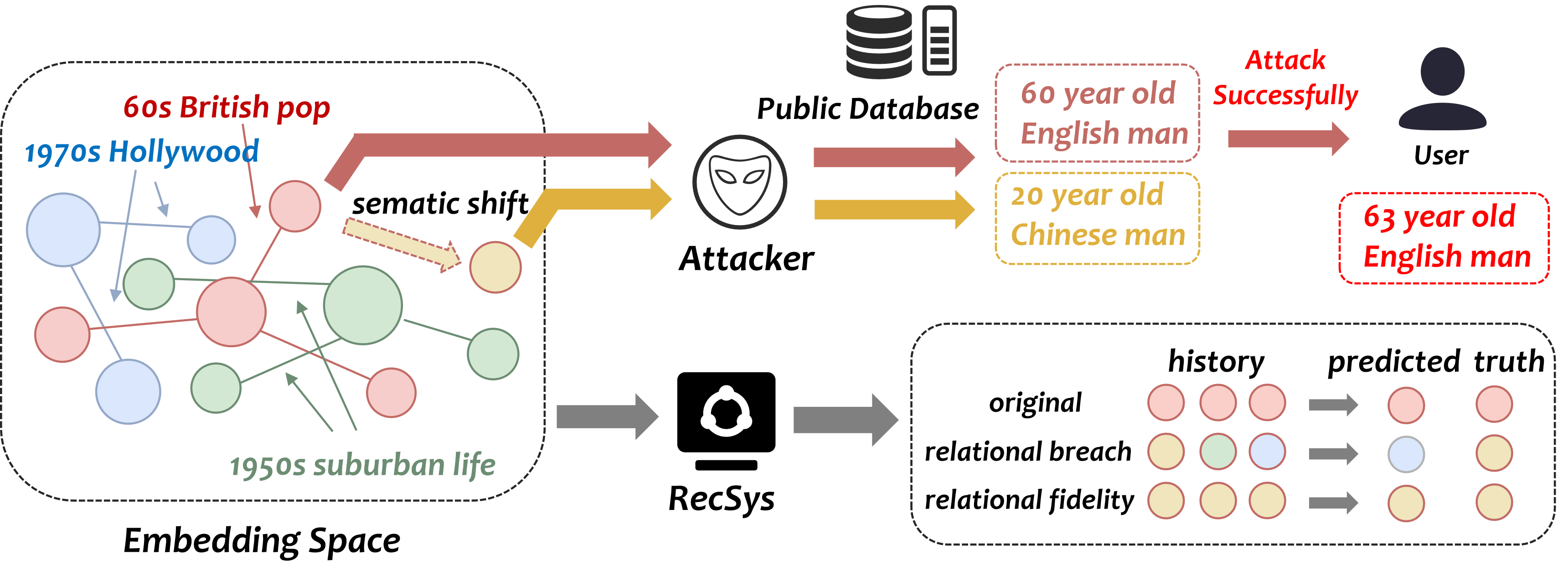}
    \caption{Motivation for relational-fidelity obfuscation in protecting item side information for LLM-enhanced recommendation.}
    \label{fig:framework_single}
\end{figure}

Faced with this input-side threat, existing defenses struggle. Post‑hoc recommendation protections such as differential privacy in recommender systems~\citep{dwork2006calibrating,mullner2023differential} and federated learning for recommendation~\citep{mcmahan2017communication,qin2021novel} assume a trustworthy collector and can hardly prevent leakage once data leaves the device. Traditional synonym‑substitution obfuscation stays at the word level~\citep{jin2020bert}, treating items as isolated tokens and easily shattering semantics, thereby destroying semantic relationships between items anchored by side information.  

To simultaneously protect privacy and maintain utility, input-side protection should achieve two semantic-level goals. The first goal is to \textbf{induce sufficient semantic shift} so that embedding inversion attacks cannot exploit the invertible mapping between embeddings and their source texts~\citep{morris2023text,zhang2025universal}. If obfuscated embeddings remain too close to the original, attackers can still approximate the text. Therefore, embeddings must move away from the original to break this invertible connection. The second goal is to \textbf{preserve local semantic neighborhood structure} that connect cold and popular items through shared attributes~\citep{vasile2016meta,liu2023graph}. Maintaining these relative relationships preserves the semantic skeleton needed for recommendation without requiring global semantic fidelity.

Traditional discrete token-level substitution cannot finely control the magnitude or direction of changes in continuous embedding space. Such perturbations disrupt local semantic relations and tightly couple deviation strength with order preservation, making joint optimization impossible. The \textit{discrete→continuous→discrete} pipeline of LLM allows fine-grained manipulation of embeddings before decoding~\citep{naparstek2026token,zhang2023llamavae}, generating new words while decoupling deviation strength from relational structure.

However, recent LLM-based text privacy methods, e.g., EmojiPrompt~\citep{lin2025emojiprompt}, LLM-Anonymization~\citep{staab2025language}, and IncogniText~\citep{frikha2025incognitext}, all follow semantic fidelity, hiding private attributes while preserving literal meaning. This misaligns with recommender systems, which need relational fidelity, i.e., preserving the relative semantic skeleton across items rather than literal text. Under relational fidelity, obfuscated content can be arbitrary as long as a usable item similarity structure persists, turning the task from constrained masking into active topology reconstruction. LLMs naturally suit this, yet existing methods were never designed for it.

To realize \textbf{relational fidelity}, we propose a \textbf{prototype-guided progressive obfuscation (PGPO) framework} with three sequential stages: prototype-aware graph construction, prototype variant exploration, and anchor-oriented graph expansion. We first identify high-influence prototypes and build a static semantic neighbor graph to capture their structural impact. The LLM then explores suitable obfuscated variants for these prototypes, which serve as stable structural anchors. Finally, the framework propagates these anchored mappings through the static graph to cover the remaining vocabulary. This process is optimized using Group Relative Policy Optimization (GRPO)~\citep{shao2024deepseekmath} with three complementary rewards: i) \textbf{individual dissimilarity} maximizes embedding distance to enforce \emph{semantic shift}, ii) \textbf{structural consistency} preserves local similarity order to maintain the semantic distribution necessary for recommendation, and iii) \textbf{semantic anchoring} constrains generation within category-specific semantic manifolds to prevent degenerate shortcuts. By jointly optimizing these rewards, the framework balances privacy and utility, protecting sensitive information while preserving the relative semantic structure crucial for recommendation.

Our contributions are as follows:
\begin{itemize}[noitemsep, topsep=0pt]
    \item We formulate input-side obfuscation for LLM-enhanced recommendation as a topology-preserving privacy problem, introducing \textbf{relational fidelity} to defend against embedding inversion while preserving relative item semantics.
    \item We propose a \textbf{prototype-guided progressive obfuscation framework} that scales topology-aware obfuscation from high-influence prototype words to the full vocabulary.
   \item We design three GRPO rewards, i.e., individual dissimilarity, structural consistency, and semantic constraint, to jointly enforce semantic shift and preserve item relations.
    \item Under aligned CosSim, experiments on two benchmarks show that our method achieves the highest recommendation utility at comparable defense strength.
\end{itemize}

\section{Related Work}
\label{sec:related}

\nosection{LLM-Enhanced Recommender Systems}
Large language models (LLMs) have reshaped recommender systems, either as semantic enhancers enriching item metadata, user profiles, and interaction graphs~\citep{wu2024survey,wei2024llmrec}, or as direct recommenders via language modeling, prompting, instruction tuning, or parameter-efficient fine-tuning, covering frameworks like P5~\citep{geng2022recommendation}, M6-Rec~\citep{cui2022m6}, TALLRec~\citep{bao2023tallrec}, LLaRA~\citep{liao2024llara} and ReLLa~\citep{lin2024rella}, and ranking or explanation models~\citep{dai2023uncovering,li2023personalized}. However, encoding side information and interactions into continuous embeddings raises privacy concerns. User clicks can be inferred from exposure \citep{xin2023user}, and LLM recommenders are vulnerable to inversion attacks \citep{wang2025privacy}, underscoring the need for input-level privacy protection \citep{luo2024privacy}.

\nosection{Embedding Inversion Attacks}
Embedding inversion attacks aim to reconstruct original text or sensitive attributes from exposed embeddings, posing significant privacy risks in NLP. Early studies demonstrated that embedding models leak rich semantic information~\citep{song2020information,fredrikson2015model}. Subsequent work further recovered full sentences using generative decoders~\citep{li2023sentence} and inverted dense embeddings with high fidelity through iterative optimization and beam search~\citep{morris2023text}. Recent attacks further expose practical vulnerabilities under cross-model~\citep{huang2024transferable}, black-box~\citep{chen2025algen}, multilingual~\citep{chen2024text}, and zero-shot settings~\citep{zhang2025universal}, motivating input-side protection for recommendation embeddings.

\nosection{Text Obfuscation Defenses}
Text obfuscation protects privacy by perturbing raw text while preserving downstream utility. Existing approaches include word- or character-level substitution and adversarial perturbation~\citep{ebrahimi2018hotflip,jia2017adversarial}, homograph or typo-based replacement, and context-aware embedding perturbation~\citep{zhou2023textobfuscator}, as well as differential privacy-based sanitization such as SanText/SanText+~\citep{yue2021differential} and local DP token mapping~\citep{feyisetan2020privacy,qu2021natural,mai2023split}. Other methods exploit controlled paraphrasing, masking, or hybrid perturbations to hide sensitive attributes. Recent work explores LLM-based privacy protection, including prompt obfuscation~\citep{lin2025emojiprompt}, LLM-based anonymization~\citep{staab2025language}, conditional attribute randomization~\citep{frikha2025incognitext}, utility-aware rewriting~\citep{yang2025robust}, and privacy-preserving LLM inference~\citep{tong2025inferdpt}. However, these studies mostly target general NLP text, user-generated content, or LLM prompts, leaving privacy-preserving obfuscation for item-side textual metadata in recommender systems largely underexplored.

\section{Method}
\label{sec:method}

\subsection{Problem Formulation}

We study input-side privacy protection for recommender systems, where item-side textual information is obfuscated before entering the downstream recommender. Let $x\in\mathcal{W}$ denote an original side-information word, and $z\in\mathcal{W}'$ its obfuscated variant generated using the static neighbors and cached history samples of $x$. Specifically, $\mathbf{u}=\phi(x)$ is the public top-$K$ neighbor set retrieved from the static semantic graph, while $\mathcal{H}$ contains the cached original--variant pairs selected from $\mathbf{u}$ and the global recent queue. The generation process is
\begin{equation}
z \sim P(\cdot \mid x,\mathbf{u},\mathcal{H};\xi),
\end{equation}
where $\xi$ denotes stochastic decoding and candidate-exploration randomness.

We conservatively consider a white-box attacker who may obtain the locally optimized conversion model and knows the original vocabulary, static semantic graph, retrieval function $\phi$, and generation prompt. By compromising the downstream recommender, the attacker may further obtain the released obfuscated text or variant set, but not the hidden generation trajectory $\Gamma$, including the processing order, cache and queue states, sampled candidates, and original--variant alignments in $\mathcal{H}$. Let $\mathcal{O}$ denote all attacker-observable auxiliary information under the corresponding setting. The privacy objective is to maximize $H(X|Z,\mathcal{O})$, equivalently minimizing $I(X;Z|\mathcal{O})$.

We learn a mapping $f:\mathcal{W}\rightarrow\mathcal{W}'$ according to the following two criteria.

\textbf{Definition 1 (Individual Dissimilarity).} Minimize the semantic similarity between a word and its obfuscated version to defeat inversion:
\begin{equation} \begin{aligned}
\min \; \mathbb{E}_{x\sim\mathcal{W}}\big[\mathrm{sim}\big(e(x), e(f(x))\big)\big],
\end{aligned} \end{equation}
where $e(\cdot)$ is the text encoder and $\mathrm{sim}$ is cosine similarity.

\textbf{Definition 2 (Relational Consistency).}
Maximize the correlation between each word's similarities with its neighbors and those of their obfuscated counterparts, preserving
the local semantic topology:
\begin{equation}
\begin{gathered}
\mathbf{s}_i=
\left[
\mathrm{sim}
\left(
e(x_i),e(x_j)
\right)
\right]_{x_j\in\mathbf{u}_i},
\\
\widetilde{\mathbf{s}}_i=
\left[
\mathrm{sim}
\left(
e(f(x_i)),e(f(x_j))
\right)
\right]_{x_j\in\mathbf{u}_i},
\\
\max_f \quad
\mathbb{E}_{x_i\sim\mathcal{W}}
\left[
\mathrm{Corr}
\left(
\mathbf{s}_i,\widetilde{\mathbf{s}}_i
\right)
\right],
\end{gathered}
\end{equation}
with $\mathrm{Corr}$ e.g.\ Spearman's rank correlation.

\begin{figure*}[t]
    \centering
    \includegraphics[width=\textwidth]{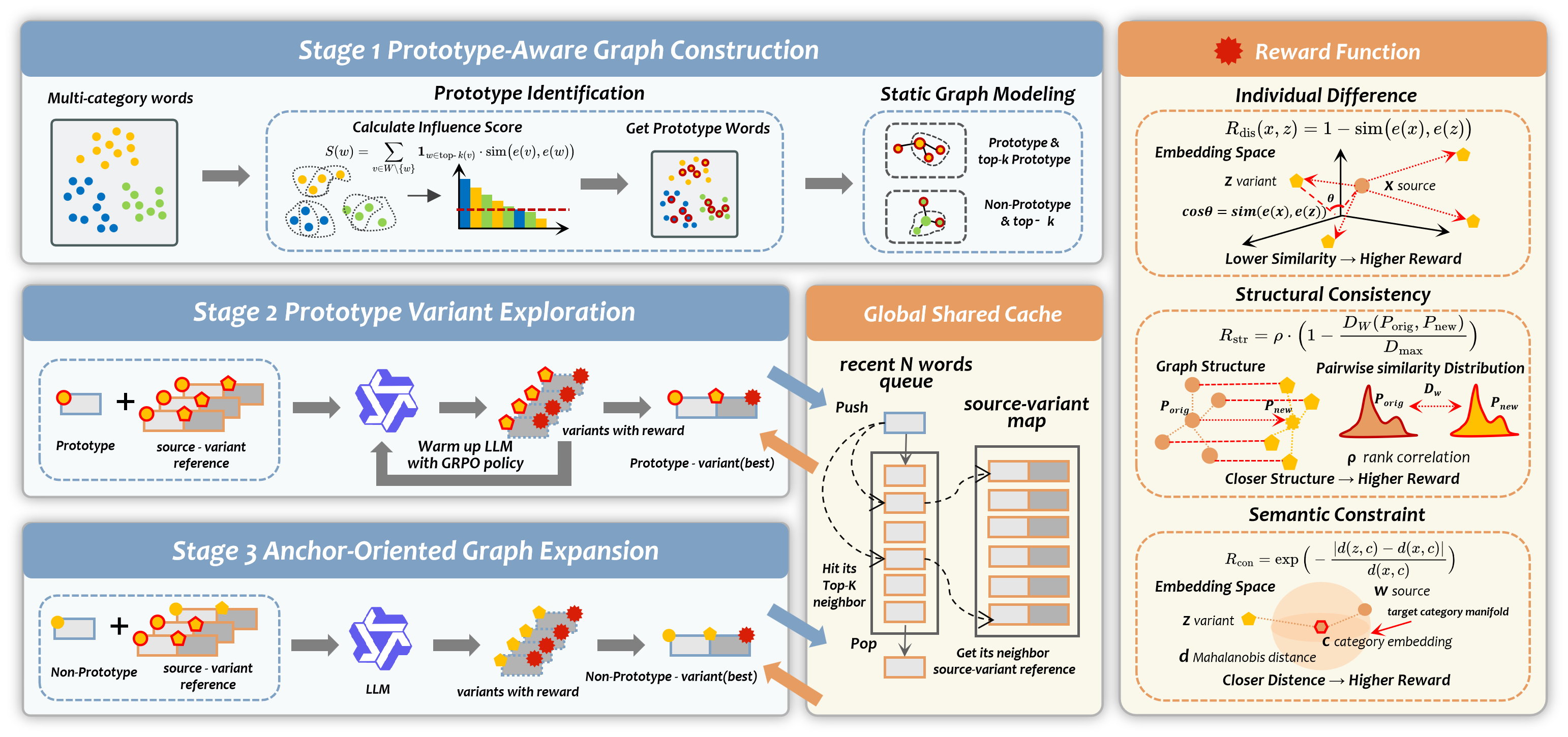}
    \caption{Overview of PGPO. High-influence prototype words are first identified, and a vocabulary-level static semantic graph is constructed. The LLM then explores prototype obfuscation anchors under GRPO-based reward optimization and propagates them through the graph to the remaining vocabulary.}
    \label{fig:framework}
\end{figure*}

We balance privacy-preserving semantic displacement against relational consistency, since stronger displacement can improve privacy but also distort local semantic structure. We therefore optimize in the continuous embedding space to better preserve relational fidelity while maintaining sufficient privacy.

\subsection{Overall Framework}
The core difficulty is that full-vocabulary joint optimization is challenging due to the large vocabulary size. We observe, however, that the global semantic topology is not uniformly distributed: a small set of \textbf{prototype words} exerts disproportionate structural influence. Based on this, we propose a \textbf{prototype-guided progressive obfuscation framework}, shown in Figure~\ref{fig:framework}. The pipeline operates in three stages:
\begin{itemize}[noitemsep, topsep=0pt]
    \setlength{\itemsep}{0pt} \setlength{\parskip}{0pt}
    \item \textbf{Identify} prototype words and build a static semantic neighbor graph.
    \item \textbf{Explore} suitable obfuscated variants for prototypes as structural anchors.
    \item \textbf{Propagate} these anchored mappings to the full vocabulary through the graph.
\end{itemize}

\subsubsection{Prototype-Aware Graph Construction}
\nosection{Motivation} The global similarity topology is shaped predominantly by a few central words. On one hand, identifying these prototypes gives us high-leverage anchor points; on the other hand, precomputing a static neighbor graph decouples expensive similarity computation from generation, enabling scalable context retrieval.

\nosection{Prototype Identification} The influence score of a word $w$ is the total similarity received from other words that rank $w$ among their top‑k neighbors:
\begin{equation} \begin{aligned}
S(w) = \sum_{v \in W \setminus \{w\}} \mathbf{1}_{w \in \operatorname{top-}k(v)} \cdot \operatorname{sim}\bigl(e(v), e(w)\bigr).
\end{aligned} \end{equation}
Sorting all words by $S(w)$ in descending order, we select the top $M$ to form the prototype set $C_{\text{Prototype}}$. The hyperparameter $k$ balances structural coverage and computational cost.

\nosection{Static Graph Modeling} For each $w \in \mathcal{W}$, we precompute and cache its top-$K$ semantic nearest neighbors based on cosine similarity. This graph later serves as an efficient index for retrieving semantically relevant context during generation.

\subsubsection{Prototype Variant Exploration}
To establish a reliable structural skeleton, we first learn obfuscation mappings for the prototype set $C_{\text{Prototype}}$ through a two-stage process.

\noindent\textbf{1) Warm-up Phase.} For $T_{\text{warmup}}$ steps, we fine-tune the LLM on $C_{\text{Prototype}}$ using GRPO. This allows the model to rapidly acquire the obfuscation behavior and, guided by the structural consistency reward, begin shaping the relative semantic relationships among prototypes.

\noindent\textbf{2) Prototype Generation Phase.} After warm-up, the model is frozen. Each prototype $w\in C_{\text{Prototype}}$ is obfuscated conditioned on its top-$K$ neighbors with cached variants from $C_{\text{Prototype}}$, producing a coherent set of prototype mappings that anchor the global topology.

%\subsubsection{Prototype-guided Graph Expansion}
\subsubsection{Anchor-Oriented Graph Expansion}
Once all prototypes are processed, the remaining words $\mathcal{W}\setminus C_{\text{Prototype}}$ are obfuscated. For each non-prototype $w$, we retrieve its top-$K$ neighbors from the static graph and include those with an already finalized variant as few-shot examples. Since high-influence prototypes appear in many neighborhoods, the structural skeleton established earlier smoothly propagates outward. This hierarchical progression from core to periphery mirrors the underlying semantic influence structure, making large-scale, topology-aware obfuscation efficient and coherent.

\subsection{Prompt Template and Caching Mechanism}
We adopt a structured prompt template with a dynamic \texttt{\{history\_samples\}} field. At generation time, each example is formatted as an original word and its finalized obfuscation.
\begin{lstlisting}
Rewrite the term "{w}" (category: {c(w)}) into one new term.
Strict rules:
- Must strictly belong to the "{c(w)}" category and sound completely natural.
- Use exactly 2 to 4 words, matching the original term's style, length, and rhythm.
- Zero overlap: Do not use any synonyms, rephrasing, similar forms, or parts of the original term.
Output exactly and only: <answer>new term</answer>
{context_str}{history samples}
\end{lstlisting}

A two-level caching system maintains stability:
\begin{itemize}[noitemsep, topsep=0pt]
    \setlength{\itemsep}{0pt} \setlength{\parskip}{0pt}
    \item \textbf{Per-word Cache:} stores the highest-reward variant from the latest group for each word.
    \item \textbf{Global Recent Queue:} tracks context-available words, with size $N$ in training and no limit in inference.
\end{itemize}

Context assembly has three steps: (i) retrieve top-$K$ static neighbors, (ii) filter by the global queue, and (iii) insert cached variants into the prompt.

\subsection{Reward-Guided Relational Fidelity Optimization}

We optimize the LLM with GRPO, whose group sampling enables diverse, critic-free training. To balance semantic shift and distribution preservation while avoiding degeneration, we use a weighted sum of three normalized rewards in $[0,1]$:
\begin{equation}
\label{eq:total_reward}
\begin{aligned}
R_{\text{total}} = \lambda_{\mathrm{dis}} R_{\text{dis}} + \lambda_{\mathrm{str}} R_{\text{str}} + \lambda_{\mathrm{con}} R_{\text{con}}.
\end{aligned}
\end{equation}
\nosection{Individual Dissimilarity Reward} 
This reward operationalizes Criterion 1. It penalizes semantic proximity between the obfuscated output and the original word, pushing the embedding beyond the attacker's local reconstruction range:
\begin{equation}
\label{eq:reward_dis}
\begin{aligned}
R_{\text{dis}}(x,z) = 1 - \mathrm{sim}\big(e(x), e(z)\big).
\end{aligned}
\end{equation}
\nosection{Structural Consistency Reward} 
This reward operationalizes Criterion 2 by evaluating whether the local similarity ordering of context words is preserved after obfuscation. Given a set of context pairs $\mathcal{H} = \{(x_i, z_i)\}_{i=1}^{|\mathcal{H}|}$, we compute pairwise similarity distributions:
\begin{equation}
\begin{aligned}
P_{\mathrm{orig}}(j)
=
\frac{\mathrm{sim}(x,x_j)+1}
{\sum_l (\mathrm{sim}(x,x_l)+1+\epsilon)},\\
P_{\mathrm{new}}(j)
=
\frac{\mathrm{sim}(z,z_j)+1}
{\sum_l (\mathrm{sim}(z,z_l)+1+\epsilon)},
\end{aligned}
\end{equation}
where $\epsilon=2\times10^{-10}$ is a numerical stability constant. Let $D_W$ denote the Wasserstein distance between these distributions~\citep{villani2009optimal}, and $\rho$ the Spearman correlation of the raw similarity matrices~\citep{spearman1961proof}. The reward is
\begin{equation}
\label{eq:reward_str}
\begin{aligned}
R_{\text{str}} = \rho \cdot \Big(1 - \frac{D_W(P_{\text{orig}}, P_{\text{new}})}{D_{\max}}\Big), 
\end{aligned}
\end{equation}
where $D_{\max}=0.5$ is empirically determined. This dual-term reward encourages both the preservation of similarity order (via $\rho$) and the overall distribution (via $D_W$).

\nosection{Semantic Constraint Reward} 
While the previous rewards encourage semantic shift and relational preservation, the lack of spatial constraints may result in degenerate outputs. We therefore introduce a semantic constraint reward to constrain obfuscation within the category-specific embedding manifold. The semantic constraint reward is defined as
\begin{equation}
\label{eq:reward_con}
R_{\text{con}} = \exp\Big( - \frac{|d(z,c) - d(x,c)|}{d(x,c)} \Big),
\end{equation}
where $d(z,c)$ denotes the Mahalanobis distance of embedding $z$ to the category center $\mu_c$ with covariance $\Sigma_c$~\citep{weinberger2009distance}.

\nosection{Training and Inference}  
During training, $R_{\text{total}}$ drives the GRPO policy updates. During inference, the best candidate from each group is selected. Joint optimization automatically balances semantic shift and distribution preservation, maintaining the relational fidelity essential for recommendation.

\section{Theoretical Analysis}
\label{sec:theory}

Detailed derivations are deferred to Appendix~\ref{app:theoretical_analysis}. 
Let $\Gamma$ be the residual hidden generation trajectory and $\mathcal{O}$ the attacker-observable information.

\begin{lemma}[Trajectory Leakage Decomposition]
\label{lem:traj}
For the obfuscated output $Z\sim p(\cdot|X,\mathcal{O},\Gamma)$,
\begin{align}
I(X;Z|\mathcal{O})=
&\ H(Z|\mathcal{O})
- \mathbb{E}\big[H(Z|X,\mathcal{O},\Gamma)\big] \notag \\
&- I(Z;\Gamma|X,\mathcal{O}).
\label{eq:traj_decomp}
\end{align}
\end{lemma}

\begin{lemma}[Semantic-Mode Entropy]
\label{lem:mode}
Partition the output space into semantic modes with weights $\alpha_m$ and intra-mode distributions $p_m$. Then
\begin{align}
\mathbb{E}\big[H(Z|X,\mathcal{O},\Gamma)\big]
=
&\ \mathbb{E}\!\left[
H(\alpha)+\sum_m \alpha_m H(p_m)
\right] \notag \\
&-\epsilon ,
\label{eq:mode_entropy}
\end{align}
where $\epsilon=\mathbb{E}[H(M|Z,X,\mathcal{O},\Gamma)]$ measures residual mode ambiguity.
\end{lemma}

With the reward-shaped policy $p_{\lambda}$ induced by 
%$R_{\mathrm{total}}=\lambda_{\mathrm{dis}}R_{\mathrm{dis}}+\lambda_{\mathrm{str}}R_{\mathrm{str}}+\lambda_{\mathrm{con}}R_{\mathrm{con}}$,
Eq.~\eqref{eq:total_reward},
these lemmas lead to the main privacy guarantee.

\begin{theorem}[Attack Success Bound for PGPO]
\label{thm:bound}
For a finite original-word set $\mathcal{X}$ and an approximately uniform prior,

{\small
\begin{align}
P_{\mathrm{succ}}
\le
\frac{1}{\log_2|\mathcal{X}|}
\Bigg(
& \epsilon^{\lambda}
- \mathbb{E}_{\lambda}\!\left[
H(\alpha^{\lambda})
+\sum_m \alpha_m^{\lambda}H(p_m^{\lambda})
\right] \notag \\
&+ H_{\lambda}(Z|\mathcal{O})
- I_{\lambda}(Z;\Gamma|X,\mathcal{O})
+ 1
\Bigg).
\label{eq:attack_bound}
\end{align}}
Here $R_{\mathrm{dis}}$ promotes mode-weight entropy $H(\alpha^{\lambda})$, $R_{\mathrm{con}}$ preserves valid intra-mode uncertainty, $R_{\mathrm{str}}$ controls $\epsilon^{\lambda}$, and $I_{\lambda}(Z;\Gamma|X,\mathcal{O})$ captures trajectory-induced uncertainty. 
\end{theorem}

For the final item representation $\mathbf{h}_{\mathrm{item}}$, data processing gives
\begin{equation}
\label{eq:rep_dpi}
I(X;\mathbf{h}_{\mathrm{item}}|C,\mathcal{O})
\le
I(X;e(Z),S|C,\mathcal{O}).
\end{equation}
If collaborative signals contain limited residual information about $X$, we further have
\begin{equation}
\label{eq:rep_bound}
I(X;\mathbf{h}_{\mathrm{item}}|C,\mathcal{O})
\lesssim
\rho_\alpha I(X;e(Z)|C,\mathcal{O})+\eta+\delta,
\end{equation}
suggesting that reducing text-level leakage also improves representation-level privacy.

\section{Experiments}
\label{sec:experiments}
\subsection{Experimental Setup}
\nosection{Datasets and Side Information}
We evaluate on two public recommendation benchmarks, MovieLens-1M\footnote{\url{https://grouplens.org/datasets/movielens/1m/}} and Amazon-Book\footnote{\url{https://jmcauley.ucsd.edu/data/amazon_v2/}}. 
Ratings no less than 3 are treated as positive for MovieLens-1M, and ratings no less than 4 for Amazon-Book. 
We use Qwen2.5-14B~\citep{Yang2024Qwen25TR} to generate multi-category item attributes from original titles and construct the side information vocabulary $\mathcal{W}$. 
Dataset statistics are shown in Table~\ref{tab:dataset}, with preprocessing and attribute construction details in Appendix~\ref{app:dataset_preprocessing} and Appendix~\ref{app:attribute_construction}.

\nosection{Recommendation Models}
We evaluate six representative recommenders: LightGCN~\citep{he2020lightgcn}, SASRec~\citep{kang2018self}, FREEDOM~\citep{zhou2023tale}, LightGT~\citep{wei2023lightgt}, TALLRec~\citep{bao2023tallrec}, and LLaRA~\citep{liao2024llara}. 
%Traditional and multimodal models are evaluated by HR@10 and NDCG@10, while LLM-based models follow their original protocols. 
More details are provided in Appendix~\ref{app:recommendation_metrics}.

\nosection{Baselines}
We compare with six baselines: None, raw Description, InferDPT~\citep{tong2025inferdpt}, Text2Text~\citep{feyisetan2020privacy,qu2021natural}, SanText+~\citep{yue2021differential}, and TextObfuscator~\citep{zhou2023textobfuscator}. 
Detailed baseline descriptions are provided in Appendix~\ref{app:baselines}.

\begin{table}[t]
\centering
\small
\renewcommand{\arraystretch}{0.95}

\begin{tabular*}{\columnwidth}{
@{\hspace{12pt}\extracolsep{\fill}}
c c c
@{\hspace{12pt}}
}

\toprule
\textbf{Dataset}
& \textbf{MovieLens-1M}
& \textbf{Amazon-Book} \\
\midrule
Users        & 6,038   & 5,921   \\
Items        & 3,629   & 5,658   \\
Interactions & 337,228 & 101,305 \\
Sparsity     & 98.46\% & 99.66\% \\
\bottomrule
\end{tabular*}

\caption{Dataset statistics.}
\label{tab:dataset}
\end{table}

\begin{table*}[t]
\centering
\footnotesize
\renewcommand{\arraystretch}{1.0}
\setlength{\tabcolsep}{6.5pt}
\setlength{\arrayrulewidth}{\cmidrulewidth}
\begin{adjustbox}{max width=\textwidth}
\begin{tabular}{cccccccccc}
\toprule

\textbf{Dataset}
& \textbf{Method}
& \textbf{Valid$^{\dagger}$}
& \textbf{CosSim$^{\dagger}$}
& \textbf{KL$^{\ddagger}$}
& \textbf{Spearman$^{\dagger}$}
& \textbf{NLL$^{\ddagger}$}
& \textbf{D-2$^{\ddagger}$}
& \textbf{D-1$^{\dagger}$}
& \textbf{LR$^{\dagger}$}
\\

\midrule

% ==================== MovieLens-1M ====================

\cellcolor{gray!18}
& Description
& -- & 1.0000 & -- & 1.0000
& 1.9858 & 0.3958 & 0.1959 & 1.0000
\\

\cellcolor{gray!18}
& InferDPT
& 0.8276 & 0.6378 & 6.7429 & 0.0546
& 61.7860 & 0.5677 & 0.2708 & 0.7379
\\

\cellcolor{gray!18}
& Text2Text
& 0.9659 & 0.6246 & 6.1386 & 0.0857
& 39.8507 & 0.4446 & 0.3661 & 0.9674
\\

\cellcolor{gray!18}
& SanText+
& 0.9912 & 0.6266 & 8.7132 & 0.1612
& 3.5427 & 0.2668 & 0.1118 & 1.2536
\\

\cellcolor{gray!18}
& TOF
& 0.1000 & 0.6297 & 6.2628 & 0.0341
& 42.7631 & 0.5455 & 0.2585 & 1.0944
\\

\cellcolor{gray!18}
\multirow{-6}{*}{\textbf{MovieLens-1M}}
& \ourscell{\textbf{Ours}}
& \ourscell{0.9986}
& \ourscell{0.6293}
& \ourscell{7.9264}
& \ourscell{0.1981}
& \ourscell{5.7050}
& \ourscell{0.4257}
& \ourscell{0.1960}
& \ourscell{0.8758}
\\

\midrule

% ==================== Amazon-Book ====================

\cellcolor{gray!18}
& Description
& -- & 1.0000 & -- & 1.0000
& 2.9403 & 0.3310 & 0.1539 & 1.0000
\\

\cellcolor{gray!18}
& InferDPT
& 0.8038 & 0.6762 & 6.2585 & 0.0538
& 50.2972 & 0.4773 & 0.2466 & 0.7777
\\

\cellcolor{gray!18}
& Text2Text
& 0.9719 & 0.6509 & 6.3891 & 0.0951
& 34.3471 & 0.3698 & 0.3242 & 0.9343
\\

\cellcolor{gray!18}
& SanText+
& 0.9899 & 0.6455 & 8.9465 & 0.1572
& 3.9354 & 0.2258 & 0.0992 & 0.9585
\\

\cellcolor{gray!18}
& TOF
& 0.9999 & 0.6668 & 6.2941 & 0.0359
& 42.5229 & 0.4624 & 0.2277 & 1.0812
\\

\cellcolor{gray!18}
\multirow{-6}{*}{\textbf{Amazon-Book}}
& \ourscell{\textbf{Ours}}
& \ourscell{0.9983}
& \ourscell{0.6519}
& \ourscell{6.9642}
& \ourscell{0.1808}
& \ourscell{10.0452}
& \ourscell{0.3378}
& \ourscell{0.1661}
& \ourscell{0.9628}
\\

\bottomrule
\end{tabular}
\end{adjustbox}

\vspace{2pt}
\parbox{\textwidth}{%
\footnotesize
$^{\dagger}$ Word-level metrics;
$^{\ddagger}$ item-description-level metrics.
}

\caption{
Quality evaluation of obfuscated variants on MovieLens-1M and
Amazon-Book.
}
\label{tab:word_combined}
\end{table*}

\begin{table*}[t]

\centering
\footnotesize
\renewcommand{\arraystretch}{1.0}
\setlength{\tabcolsep}{2.6pt}
\begin{adjustbox}{max width=\textwidth}
\begin{tabular}{ccccccccccc}
\toprule

\multirow{2}{*}{\textbf{Method}}
& \multicolumn{2}{c}{\textbf{SASRec}}
& \multicolumn{2}{c}{\textbf{LightGCN}}
& \multicolumn{2}{c}{\textbf{FREEDOM}}
& \multicolumn{2}{c}{\textbf{LightGT}}
& \textbf{TALLRec}
& \textbf{LLaRA}
\\

\cmidrule(lr){2-3}
\cmidrule(lr){4-5}
\cmidrule(lr){6-7}
\cmidrule(lr){8-9}
\cmidrule(lr){10-10}
\cmidrule(lr){11-11}

& \textbf{HR@10}
& \textbf{NDCG@10}
& \textbf{HR@10}
& \textbf{NDCG@10}
& \textbf{HR@10}
& \textbf{NDCG@10}
& \textbf{HR@10}
& \textbf{NDCG@10}
& \textbf{AUC}
& \textbf{HR@1}
\\

\midrule

% ==================== MovieLens-1M ====================
\rowcolor{gray!18}
\multicolumn{11}{c}{\textbf{MovieLens-1M}}
\\

None
& \result{0.1114}{0.0012}
& \result{0.0513}{0.0005}
& \result{0.0826}{0.0007}
& \result{0.0200}{0.0008}
& \result{0.0802}{0.0019}
& \result{0.0191}{0.0008}
& \result{0.0816}{0.0011}
& \result{0.0200}{0.0003}
& \result{0.6834}{0.0027}
& \result{0.2231}{0.0123}
\\

Description
& \result{0.1129}{0.0008}
& \result{0.0523}{0.0003}
& \result{0.0850}{0.0009}
& \result{0.0204}{0.0004}
& \result{0.0840}{0.0012}
& \result{0.0207}{0.0003}
& \result{0.0847}{0.0013}
& \result{0.0206}{0.0006}
& \result{0.6990}{0.0036}
& \result{0.2454}{0.0064}
\\

\cmidrule{1-11}

InferDPT
& \result{0.1085}{0.0037}
& \result{0.0506}{0.0014}
& \result{0.0828}{0.0005}
& \result{0.0204}{0.0004}
& \second{0.0830}{0.0015}
& \result{0.0196}{0.0006}
& \result{0.0824}{0.0004}
& \result{0.0203}{0.0002}
& \result{0.6925}{0.0065}
& \result{0.2337}{0.0055}
\\

Text2Text
& \result{0.1118}{0.0017}
& \result{0.0516}{0.0007}
& \second{0.0835}{0.0006}
& \result{0.0203}{0.0004}
& \result{0.0823}{0.0011}
& \best{0.0199}{0.0005}
& \second{0.0825}{0.0019}
& \second{0.0204}{0.0004}
& \second{0.6961}{0.0027}
& \result{0.2364}{0.0150}
\\

SanText+
& \result{0.1128}{0.0011}
& \second{0.0521}{0.0004}
& \result{0.0825}{0.0010}
& \second{0.0205}{0.0004}
& \result{0.0819}{0.0010}
& \result{0.0195}{0.0002}
& \result{0.0812}{0.0011}
& \result{0.0203}{0.0002}
& \result{0.6960}{0.0019}
& \result{0.2358}{0.0074}
\\

TOF
& \second{0.1128}{0.0007}
& \result{0.0518}{0.0003}
& \result{0.0834}{0.0005}
& \result{0.0190}{0.0011}
& \result{0.0827}{0.0007}
& \result{0.0194}{0.0005}
& \result{0.0821}{0.0003}
& \result{0.0203}{0.0001}
& \result{0.6928}{0.0057}
& \second{0.2366}{0.0053}
\\

\rowcolor{cyan!9}
\textbf{Ours}
& \best{0.1133}{0.0011}
& \best{0.0528}{0.0003}
& \best{0.0848}{0.0014}
& \best{0.0207}{0.0006}
& \best{0.0842}{0.0020}
& \second{0.0197}{0.0006}
& \best{0.0839}{0.0006}
& \best{0.0208}{0.0002}
& \best{0.6978}{0.0046}
& \best{0.2451}{0.0048}
\\

\midrule

% ==================== Amazon-Book ====================
\rowcolor{gray!18}
\multicolumn{11}{c}{\textbf{Amazon-Book}}
\\

None
& \result{0.1809}{0.0024}
& \result{0.0882}{0.0009}
& \result{0.0883}{0.0040}
& \result{0.0398}{0.0024}
& \result{0.0733}{0.0017}
& \result{0.0316}{0.0009}
& \result{0.0900}{0.0016}
& \result{0.0396}{0.0006}
& \result{0.7857}{0.0128}
& \result{0.4614}{0.0113}
\\

Description
& \result{0.1843}{0.0018}
& \result{0.0905}{0.0010}
& \result{0.0926}{0.0014}
& \result{0.0419}{0.0010}
& \result{0.0762}{0.0018}
& \result{0.0328}{0.0009}
& \result{0.0929}{0.0016}
& \result{0.0413}{0.0005}
& \result{0.7983}{0.0014}
& \result{0.5348}{0.0114}
\\

\cmidrule{1-11}

InferDPT
& \result{0.1798}{0.0028}
& \result{0.0879}{0.0011}
& \result{0.0906}{0.0011}
& \result{0.0405}{0.0007}
& \second{0.0744}{0.0012}
& \second{0.0321}{0.0008}
& \result{0.0912}{0.0017}
& \result{0.0405}{0.0010}
& \result{0.7965}{0.0024}
& \second{0.5199}{0.0079}
\\

Text2Text
& \result{0.1797}{0.0022}
& \result{0.0888}{0.0014}
& \result{0.0909}{0.0014}
& \result{0.0411}{0.0010}
& \result{0.0731}{0.0018}
& \result{0.0315}{0.0011}
& \second{0.0919}{0.0017}
& \second{0.0407}{0.0012}
& \result{0.7937}{0.0037}
& \result{0.5147}{0.0159}
\\

SanText+
& \result{0.1811}{0.0030}
& \result{0.0889}{0.0006}
& \second{0.0917}{0.0019}
& \result{0.0406}{0.0007}
& \result{0.0738}{0.0015}
& \result{0.0315}{0.0010}
& \result{0.0916}{0.0024}
& \result{0.0405}{0.0011}
& \result{0.7947}{0.0044}
& \result{0.5070}{0.0225}
\\

TOF
& \second{0.1827}{0.0015}
& \second{0.0896}{0.0007}
& \result{0.0914}{0.0014}
& \second{0.0413}{0.0006}
& \result{0.0735}{0.0018}
& \result{0.0316}{0.0014}
& \result{0.0916}{0.0011}
& \result{0.0402}{0.0007}
& \second{0.7976}{0.0034}
& \result{0.5156}{0.0139}
\\

\rowcolor{cyan!9}
\textbf{Ours}
& \best{0.1846}{0.0025}
& \best{0.0903}{0.0009}
& \best{0.0920}{0.0013}
& \best{0.0414}{0.0008}
& \best{0.0753}{0.0025}
& \best{0.0326}{0.0007}
& \best{0.0930}{0.0018}
& \best{0.0412}{0.0012}
& \best{0.7979}{0.0049}
& \best{0.5257}{0.0092}
\\

\bottomrule
\end{tabular}
\end{adjustbox}
\caption{
Recommendation performance on MovieLens-1M and Amazon-Book.
}
\label{tab:rec_combined}
\end{table*}

\nosection{Privacy Evaluation}
We use Vec2Text~\citep{morris2023text} as the embedding inversion attacker to recover sensitive category labels from item embeddings~\citep{fredrikson2015model}. For fair comparison, we align the average cosine similarity between original and obfuscated embeddings across methods, and evaluate attack performance under the same aligned CosSim~\citep{morris2023text,li2023sentence,huang2024transferable}. To evaluate privacy protection under the white-box threat model, we additionally employ maximum-likelihood inversion attacker~\citep{tong2025vulnerability} and report ReID@1, MRR, $H_{\mathrm{att}}$, and $I_{\mathrm{att}}$. Detailed attack settings are provided in Appendix~\ref{app:attack_model}.

\nosection{Implementation Details}
Texts are encoded with BERT-base-uncased~\citep{devlin2019bert}. For traditional recommenders, embeddings are reduced to 128 dimensions by PCA~\citep{abdi2010principal}; for multimodal and LLM-based models, 768-dimensional embeddings are retained. Feature processing details, default hyperparameters, and reward schedules are reported in Appendix~\ref{app:feature_processing} and Appendix~\ref{app:training_configuration}.

\subsection{Overall Performance}
\label{sec:exp-rq}

We evaluate our method across three key dimensions: obfuscated variants quality, recommendation performance, and embedding inversion defense.

\nosection{(i) Obfuscated Variants Quality} As shown in Table~\ref{tab:word_combined}, our method produces valid variants for nearly all words and achieves the highest Spearman correlation on both datasets (0.1981 on MovieLens-1M, 0.1808 on Amazon-Book), substantially outperforming all baselines in preserving local semantic topology. In contrast, SanText+ yields low NLL but suffers from reduced diversity due to frequent special tokens, while InferDPT, Text2Text, and TOF generate noisy outputs with weak neighborhood consistency. These results confirm that our approach better balances obfuscation, fluency, and relational semantic preservation under comparable perturbation strength.

\begin{figure}[t]
\centering
\begin{subfigure}{0.48\linewidth}
    \centering
    \includegraphics[width=\linewidth]{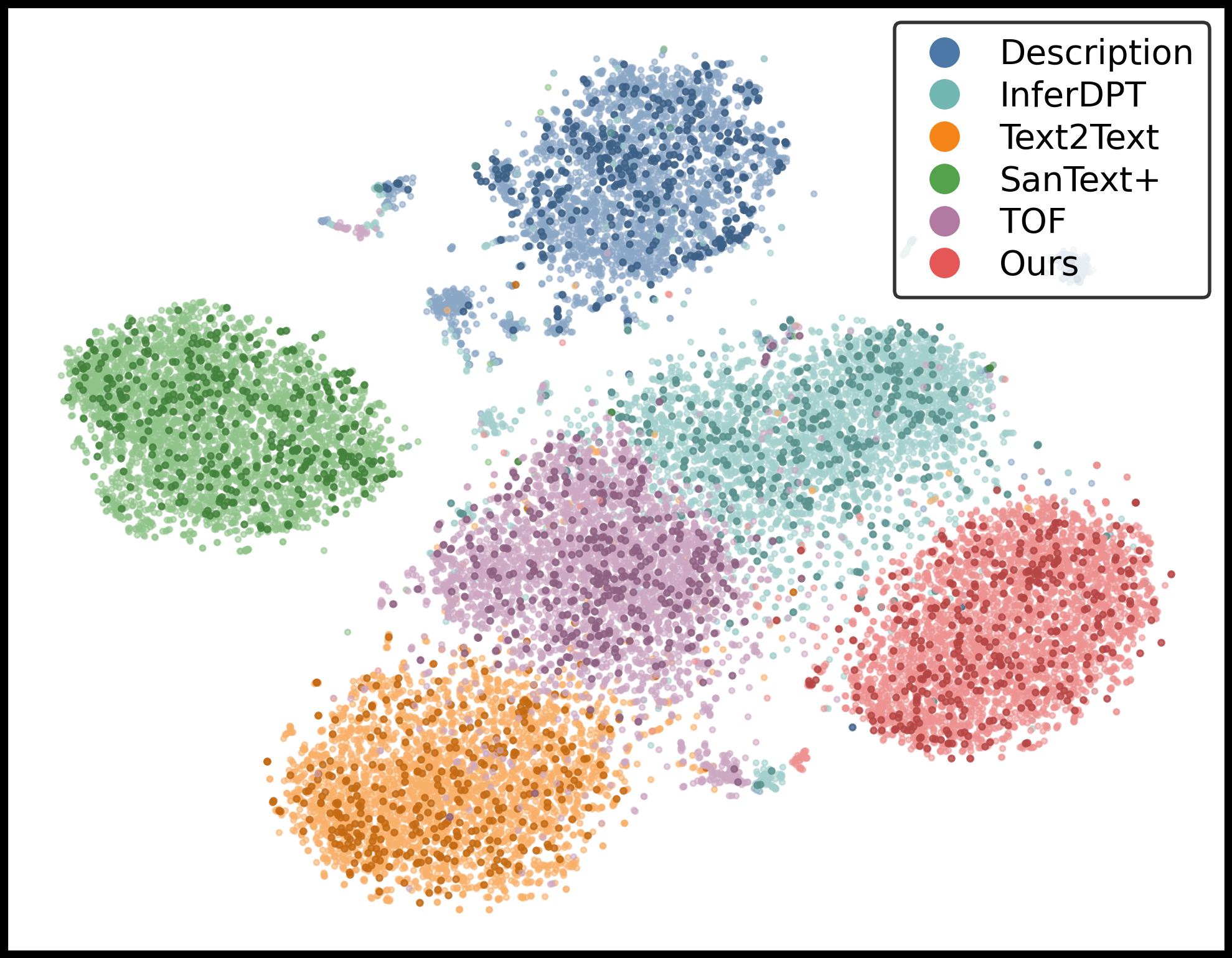}
    \caption{MovieLens-1M}
    \label{fig:sub2}
\end{subfigure}
\hfill
\begin{subfigure}{0.48\linewidth}
    \centering
    \includegraphics[width=\linewidth]{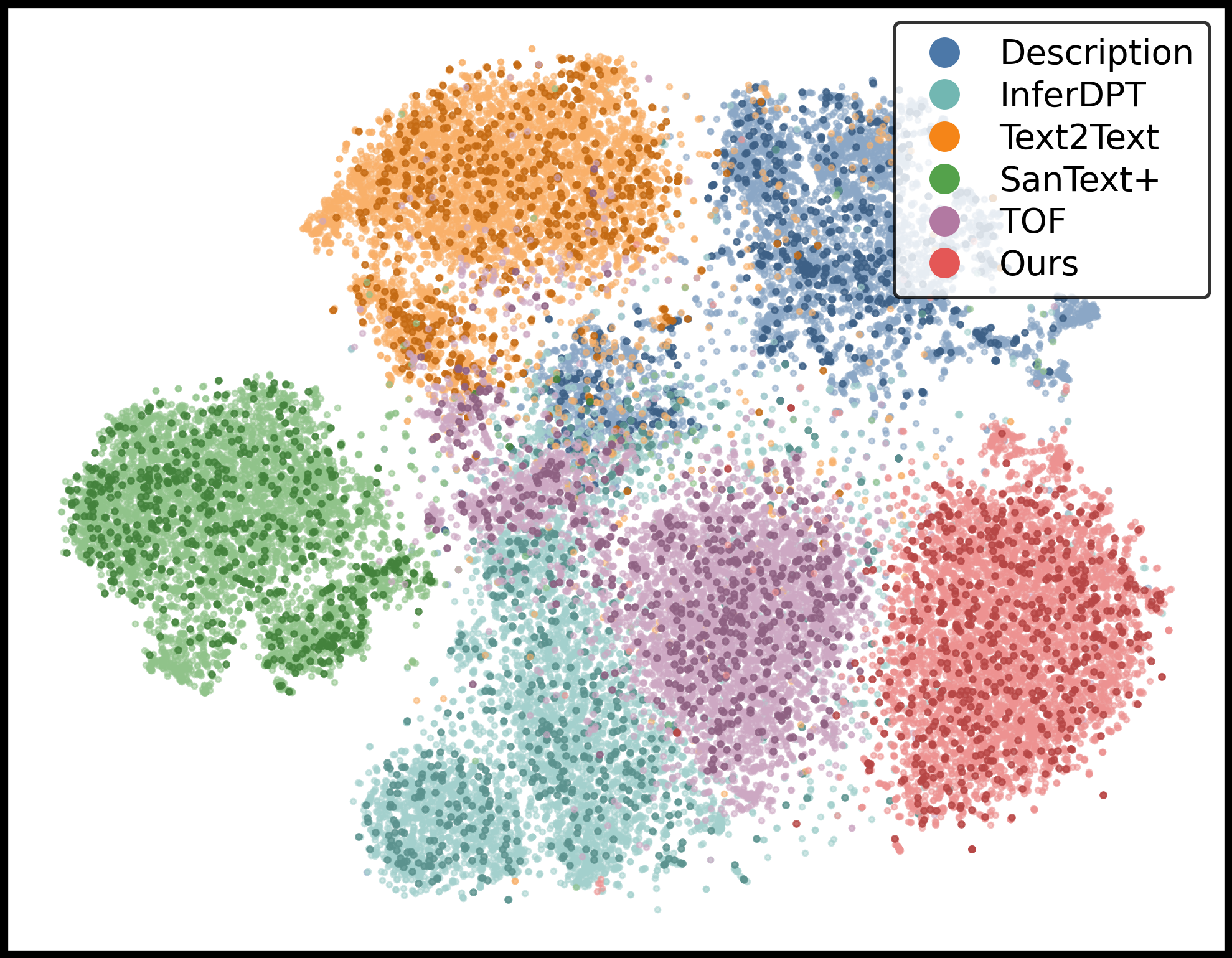}
    \caption{Amazon-Book}
    \label{fig:sub3}
\end{subfigure}

\caption{T-SNE visualization of item embeddings from original descriptions and method-specific obfuscated variants for each item.}
\label{fig:combined}
\end{figure}

\begin{table*}[t]
\centering
\footnotesize
\renewcommand{\arraystretch}{0.9}
\setlength{\tabcolsep}{6.5pt}

\begin{adjustbox}{max width=\textwidth}
\begin{tabular}{ccccccccccc}
\toprule

\multirow{2}{*}{\textbf{Method}}

& \multicolumn{5}{c}
  {\textbf{MovieLens-1M}}
& \multicolumn{5}{c}
  {\textbf{Amazon-Book}}
\\

& \textbf{EM}
& \textbf{Precision}
& \textbf{Recall}
& \textbf{F1}
& \textbf{Jaccard}
& \textbf{EM}
& \textbf{Precision}
& \textbf{Recall}
& \textbf{F1}
& \textbf{Jaccard}
\\

\midrule

Description
& 0.0641
& 0.2413
& 0.3638
& 0.2720
& 0.2116
& 0.0988
& 0.1829
& 0.1836
& 0.1832
& 0.1551
\\

\cmidrule{1-11}

InferDPT
& 0.0309
& 0.0676
& 0.0848
& 0.0703
& 0.0583
& 0.0480
& 0.0950
& 0.0955
& 0.0952
& 0.0795
\\

Text2Text
& 0.0276
& 0.0593
& 0.0808
& 0.0646
& 0.0523
& 0.0324
& 0.0727
& 0.0730
& 0.0728
& 0.0594
\\

SanText+
& 0.0276
& 0.0839
& 0.1252
& 0.0941
& 0.0711
& 0.0226
& 0.0599
& 0.0605
& 0.0601
& 0.0477
\\

TOF
& 0.0301
& 0.0672
& 0.0912
& 0.0722
& 0.0586
& 0.0384
& 0.0887
& 0.0896
& 0.0890
& 0.0722
\\

%\rowcolor{blue!9}
\rowcolor{cyan!9}
\textbf{Ours}
& 0.0309
& 0.0662
& 0.0913
& 0.0721
& 0.0593
& 0.0314
& 0.0925
& 0.0933
& 0.0928
& 0.0724
\\

\bottomrule
\end{tabular}
\end{adjustbox}

\caption{
Embedding inversion defense comparison across MovieLens-1M and
Amazon-Book.
}
\label{tab:attack_defense}
\end{table*}

\nosection{(ii) Recommendation Performance} Table~\ref{tab:rec_combined} shows that our method consistently achieves the best or near-best downstream recommendation results across all model types on both datasets. The overall ANOVA further verifies that these performance gains are statistically significant. Compared to the upper bound using raw descriptions, the performance drop is minimal, typically below 2\% for traditional collaborative filtering models, indicating that key semantic relations in the side information are largely retained. Baseline token-level methods, however, degrade embedding quality and disrupt implicit item–item relations, leading to inferior accuracy.

\nosection{(iii) Embedding Inversion Defense} Table~\ref{tab:attack_defense} reports that all obfuscation methods cut attack success rates by over 50\% relative to the raw Description baseline, confirming that semantic perturbation effectively mitigates inversion attacks. Although baseline token-level methods offer partial protection, they degrade word quality and recommendation utility, as Tables~\ref{tab:word_combined} and~\ref{tab:rec_combined} show. In contrast, our method achieves comparable defense strength with better downstream utility.

\begin{figure}[t]
\centering

\begin{subfigure}{\linewidth}
    \centering
    \includegraphics[width=\linewidth]{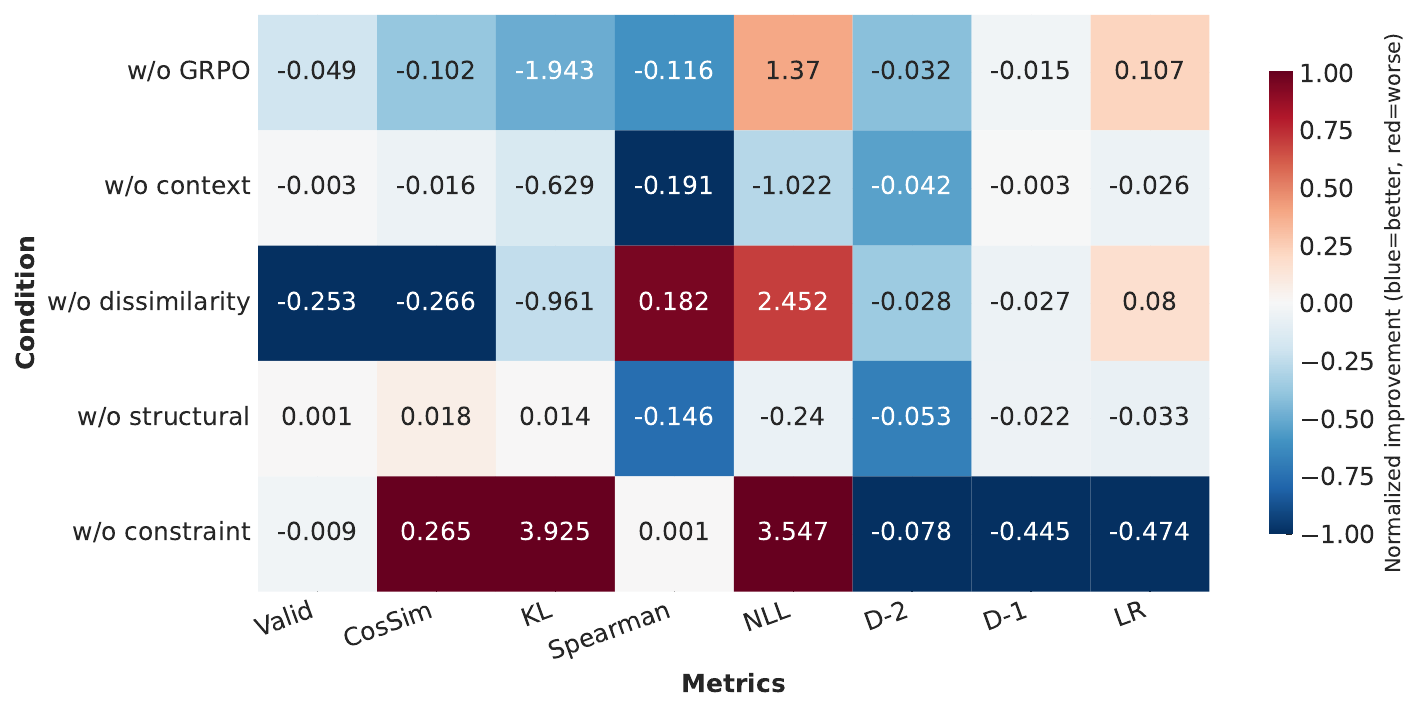}
    \caption{Normalized variant quality metric changes relative to the full model.}
    \label{fig:sub1}
\end{subfigure}

\begin{subfigure}[t]{0.48\linewidth}
    \centering
    \includegraphics[width=\linewidth]{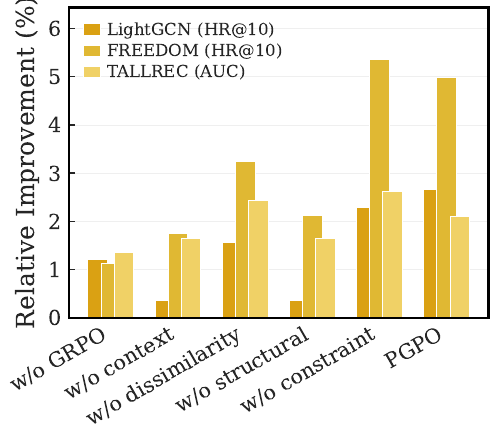}
    \caption{Recommendation utility.}
    \label{fig:sub2}
\end{subfigure}
\hfill
\begin{subfigure}[t]{0.48\linewidth}
    \centering
    \includegraphics[width=\linewidth]{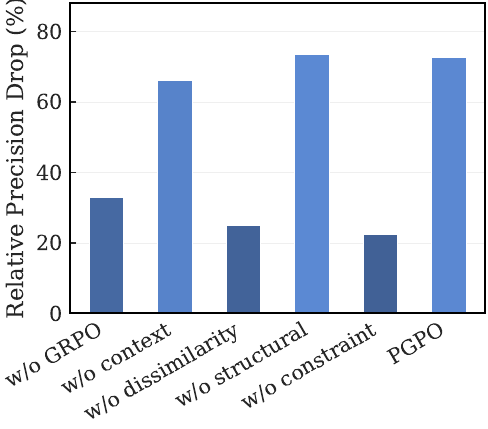}
    \caption{Inversion defence.}
    \label{fig:sub3}
\end{subfigure}
\caption{Ablation study of each component.}
\label{fig:ablation}
\end{figure}

\subsection{Ablation Study}
We conduct ablation studies to evaluate the contribution of each component, where we set variants as: (1) \textbf{w/o GRPO}, directly generating outputs with the LLM; (2) \textbf{w/o context}, removing context examples and generating each word independently; (3) \textbf{w/o dissimilarity}, removing the individual dissimilarity reward; (4) \textbf{w/o structural}, removing the structural consistency reward; and (5) \textbf{w/o constraint}, removing the semantic constraint reward.

Results in Fig.~\ref{fig:ablation} show that all components are important. The \textbf{w/o GRPO} variant reduces recommendation performance and increases attack success, confirming the benefit of group-based policy optimization. The \textbf{w/o context} variant sharply weakens semantic preservation, showing that neighborhood-aware generation is essential for maintaining local topology. The \textbf{w/o dissimilarity} variant raises CosSim, indicating insufficient semantic shift and weaker privacy protection. The \textbf{w/o structural} variant harms both semantic correlation and recommendation utility, confirming the need to preserve local semantic order. The \textbf{w/o constraint} variant degenerates into just translation, causing high attack success and semantic inconsistency, which highlights the constraint reward's role in preventing degenerate shortcuts.

\begin{table}[t]
\centering
\footnotesize
\renewcommand{\arraystretch}{0.95}

\begin{tabular*}{\columnwidth}{
@{\hspace{6pt}\extracolsep{\fill}}
c c c c c
@{\hspace{6pt}}
}
\toprule
\textbf{Method}
& \textbf{ReID@1}
& \textbf{MRR}
& $\mathbf{H_{\mathrm{att}}}$
& $\mathbf{I_{\mathrm{att}}}$ \\
\midrule
w/o GRPO           & 0.5946 & 0.6796 & 1.129 & 4.515 \\
w/o context        & 0.4406 & 0.5627 & 1.506 & 4.138 \\
w/o dissimilarity  & 0.8918 & 0.9210 & 0.268 & 5.376 \\
w/o structural     & 0.1959 & 0.3102 & 1.790 & 3.854 \\
w/o constraint     & 0.3435 & 0.4441 & 1.455 & 4.189 \\
\textbf{PGPO}      & 0.2642 & 0.3795 & 1.647 & 3.997 \\
\bottomrule
\end{tabular*}

\caption{Privacy ablation under Replay-MLE attack.}
\label{tab:replay_ablation}
\end{table}

Table~\ref{tab:replay_ablation} evaluates the information-theoretic privacy objective under a maximum-likelihood inversion attacker. The \textbf{w/o dissimilarity} variant causes the most severe privacy degradation, while the \textbf{w/o GRPO}, \textbf{w/o context}, and \textbf{w/o constraint} variants also increase re-identification risk and information leakage. Although the \textbf{w/o structural} variant strengthens original–obfuscated unlinkability, it substantially harms relational fidelity and recommendation utility.

\begin{figure}[t]
\centering

\begin{subfigure}[t]{0.48\linewidth}
    \centering
    \includegraphics[width=\linewidth]{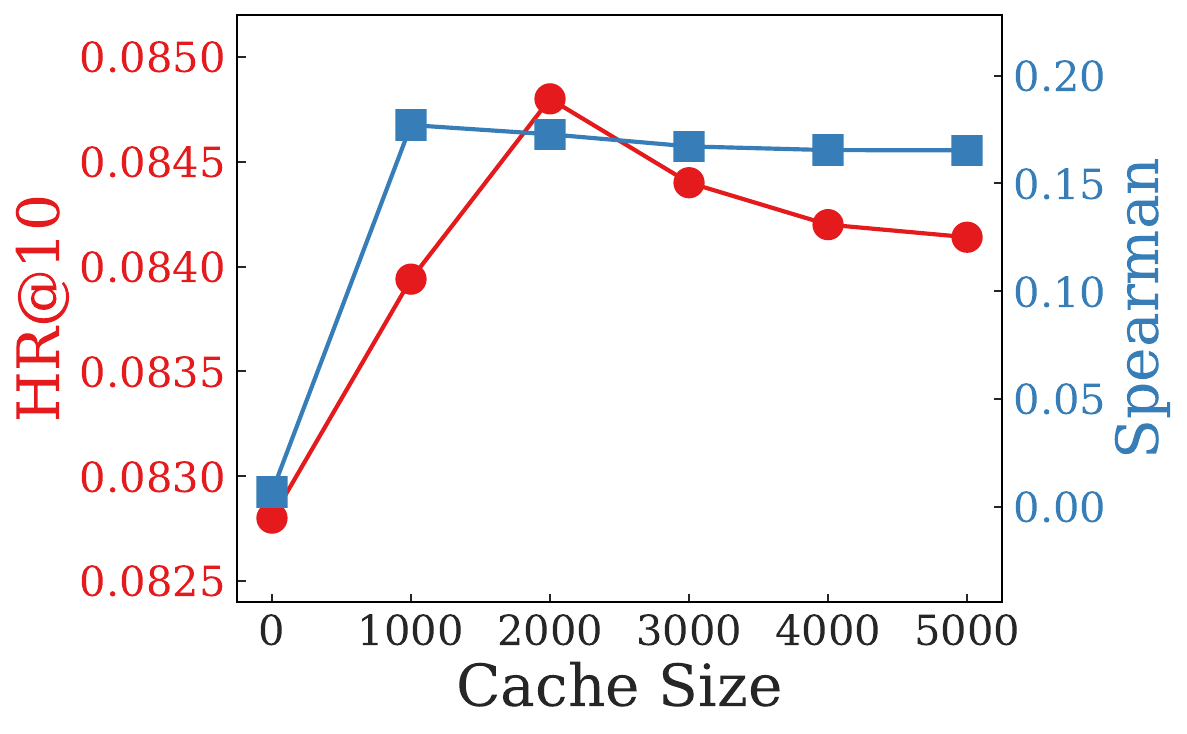}
    \caption{Recommendation utility.}
    \label{fig:psa_sub1}
\end{subfigure}
\hfill
\begin{subfigure}[t]{0.48\linewidth}
    \centering
    \includegraphics[width=\linewidth]{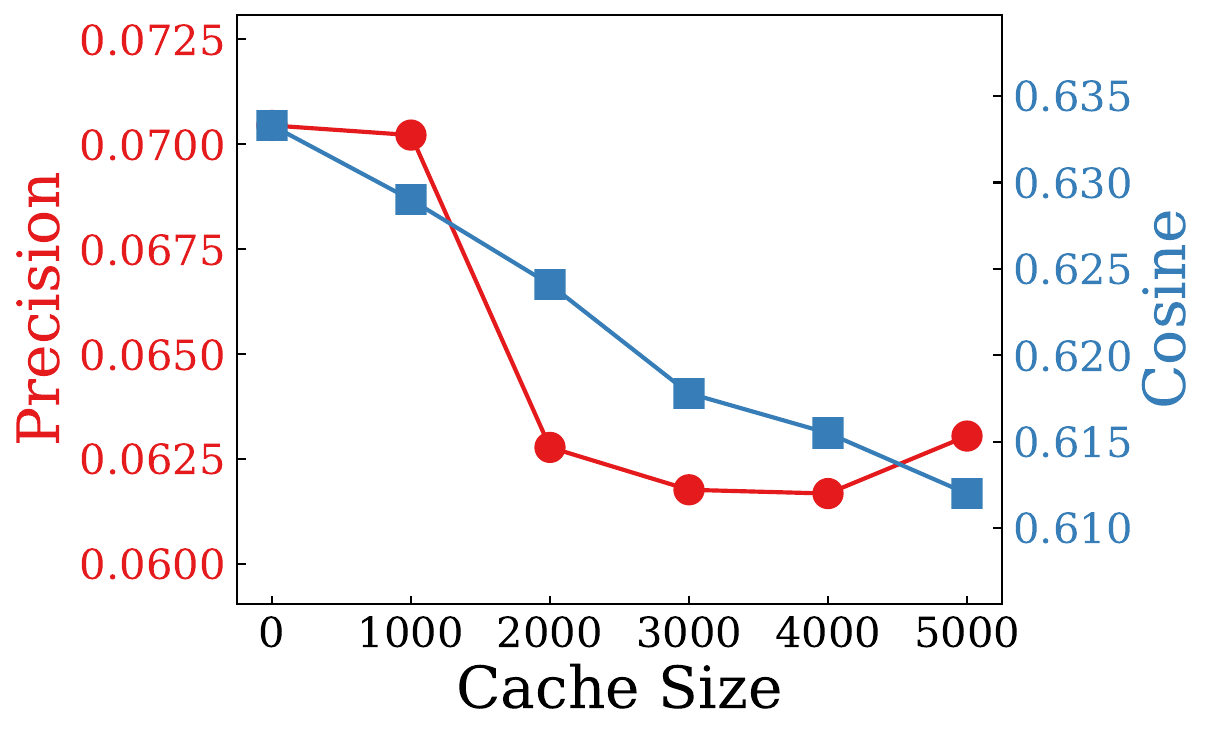}
    \caption{Inversion defence.}
    \label{fig:psa_sub2}
\end{subfigure}
\caption{Cache Size Sensitivity.}
\label{fig:psa_cache_size}
\end{figure}

\begin{figure}[t]
\centering
\begin{subfigure}[t]{0.48\linewidth}
    \centering
    \includegraphics[width=\linewidth]{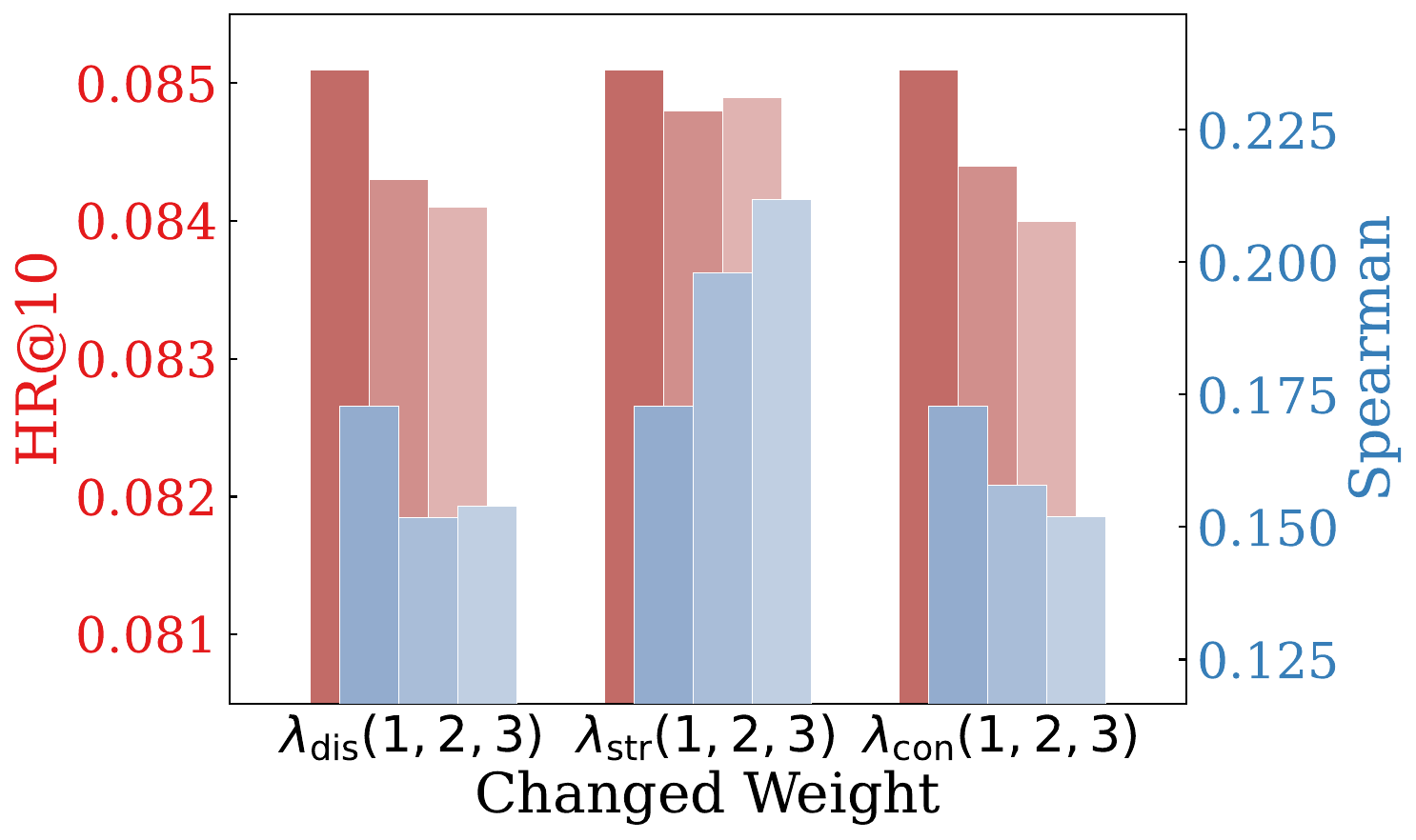}
    \caption{Recommendation utility.}
    \label{fig:psa_sub3}
\end{subfigure}
\hfill
\begin{subfigure}[t]{0.48\linewidth}
    \centering
    \includegraphics[width=\linewidth]{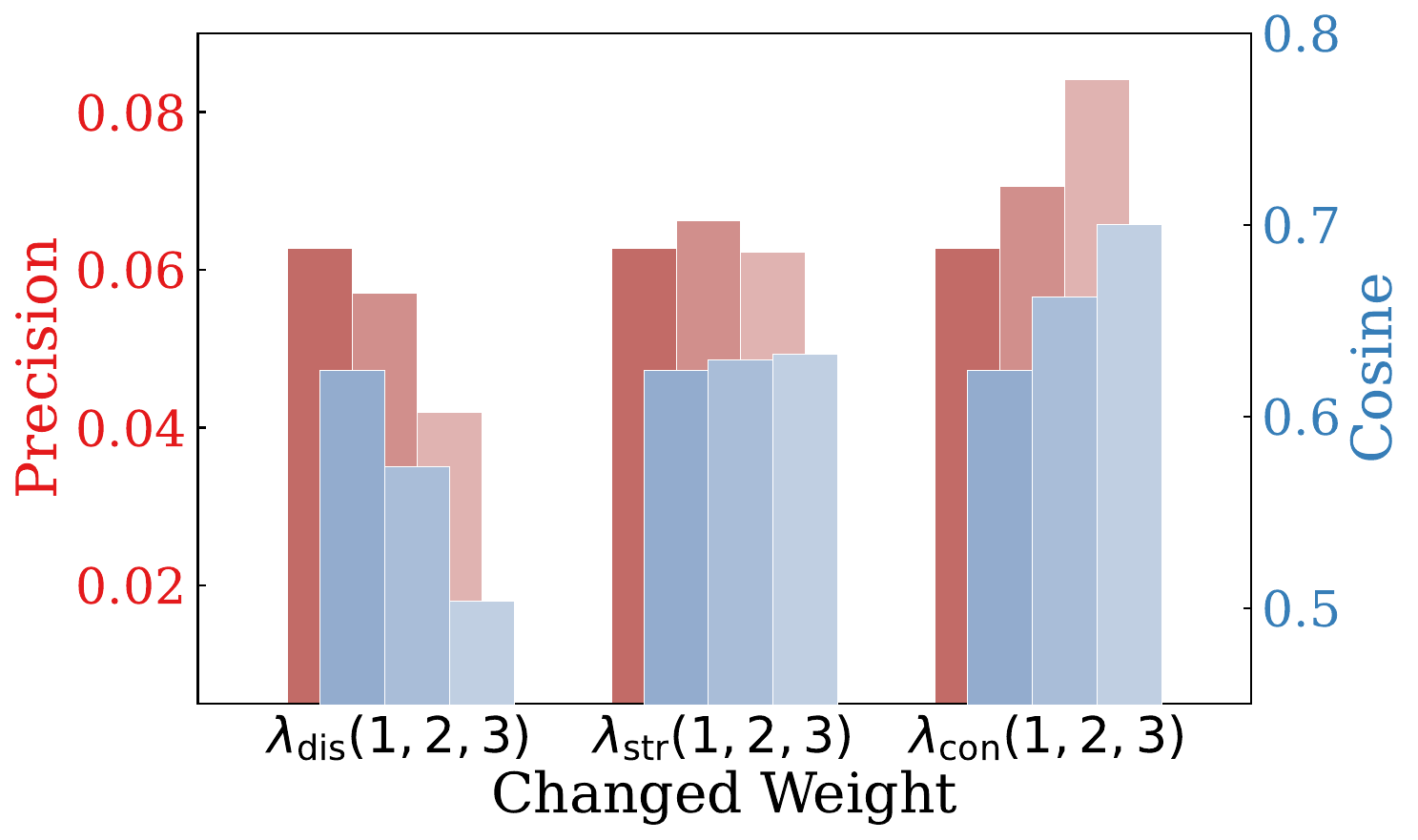}
    \caption{Inversion defence.}
    \label{fig:psa_sub4}
\end{subfigure}

\caption{Reward Weights Sensitivity.}
\label{fig:psa_weight}
\end{figure}

\subsection{Parameter Sensitivity Analysis}

\nosection{Cache Size} Figs.~\ref{fig:psa_cache_size}, \ref{fig:cache_tsne} and Table~\ref{tab:cache_psa} show that a very small cache destroys topology, a moderate cache improves Spearman correlation to preserve local semantics, and further increases slightly degrade both, indicating diminishing returns and minor perturbations from early low-quality samples.

\nosection{Prototype Word Count} Figs.~\ref{fig:core_word_tsne}, ~\ref{fig:psa_core_word} and Table~\ref{tab:coreword_psa} show few prototype words suffice to preserve semantic relations, while adding more increases structure learning difficulty, degrades performance, and biases the model toward emphasizing dissimilarity.

\nosection{Reward Weight Configuration} Figs.~\ref{fig:psa_weight}, \ref{fig:three_subplots} and Table~\ref{tab:weight_psa} show that reward weights steer the model to generate diverse variants, thereby controlling privacy strength via similarity and recommendation performance via structure preservation, with both balanced through semantic constraint.

Detailed analyses are provided in Appendix~\ref{app:additional_analysis}.

\section{Conclusion}
\label{sec:conclusion}
In this paper, we addressed input-level privacy for LLM-enhanced recommender systems. By shifting from semantic fidelity to \textbf{relational fidelity}, we proposed PGPO, a prototype-guided progressive obfuscation framework. Both theoretical analysis and experiments show that PGPO achieves a stronger privacy--utility trade-off than baselines while preserving recommendation utility. Future work may extend the framework to other modalities and decentralized deployment.

\section*{Limitations}
Our framework currently treats side-information words with a unified obfuscation strategy, without explicitly assigning higher protection priority to more sensitive or rare words over generic ones. Future work may incorporate sensitivity-aware weighting to better allocate privacy budgets across words. Additionally, PGPO introduces extra computational cost due to LLM-based variant generation, which may be mitigated in future work through offline caching, batch generation, lightweight policy distillation, or retrieval-based reuse of high-quality obfuscation mappings. Moreover, our current study focuses primarily on textual side information, leaving the extension of relational fidelity to multimodal inputs such as images, audio, video, and knowledge graphs for future work. A risk is that the obfuscation could be misused to bypass content moderation by replacing prohibited items with innocuous expressions.

% Acknowledgements (if any) can be added here as \section*{Acknowledgments}
\section*{Acknowledgments}

This work was supported by the National Natural Science Foundation 
of China under Grant Nos.~72192823, 62177038, 62522217 and 624B2131.

\bibliography{custom}

\appendix

\section{Theoretical Analysis}
\label{app:theoretical_analysis}

This appendix provides detailed derivations for Section~\ref{sec:theory}. 
Let $X$ denote the original side-information word, $\mathbf{U}=\phi(X)$ denote its public neighbor set retrieved from the static semantic graph, and $Z$ denote the generated obfuscated variant. 
We consider a strong attacker who knows the LLM parameters, the static semantic graph, the retrieval function $\phi$, and the original vocabulary. 
The attacker may also observe the released obfuscated text or the set of released variants.

We define $\mathcal{O}$ as all attacker-observable or inferable information, including the released variants, the public graph, the original vocabulary, and any final-position information if available. 
The hidden trajectory $\Gamma$ contains only residual step-wise generation factors that remain unobserved after conditioning on $\mathcal{O}$, such as unreleased cache states, queue states, sampled candidate groups, source-variant alignments inside history samples, and stochastic decoding paths. 
Thus, the privacy objective is to reduce $I(X;Z|\mathcal{O})$, equivalently increasing $H(X|Z,\mathcal{O})$.

\subsection{Path-dependent Generation Trajectory}

PGPO does not rely on hiding the released variants themselves. 
Instead, its privacy effect comes from the path-dependent generation process that determines how each prompt context is assembled and how each finalized variant is selected. 
At step $t$, let the processing order be $\pi$ and let
%\begin{equation}
$X_t = W_{\pi_t}$
%\end{equation}
be the word currently being obfuscated. 
Its static neighbor set is public and fixed by the precomputed graph:
\begin{equation}
\mathbf{U}_t = \phi(X_t).
\end{equation}

Let $\mathcal{C}_{t-1}$ denote the per-word cache before step $t$, and let $\mathcal{Q}_{t-1}$ denote the global recent queue. 
The actual history samples injected into the prompt are
\begin{equation}
\label{eq:app_history}
\mathcal{H}_t
=
\{(x_j,z_j)\in \mathcal{C}_{t-1}: x_j\in \mathbf{U}_t,\ x_j\in \mathcal{Q}_{t-1}\}.
\end{equation}
Although $\mathbf{U}_t$ is public, the realized $\mathcal{H}_t$ is path-dependent. 
Different processing orders, cache states, and queue states may cause the same word and the same static neighbor set to receive different history samples.

Given $\mathcal{H}_t$, the LLM samples a group of candidates:
\begin{equation}
\tilde{Z}_t^{(g)}
\sim
p_{\theta_t}
(\cdot|X_t,\mathbf{U}_t,\mathcal{H}_t;\xi_t),
\quad
g=1,\ldots,G,
\end{equation}
where $\xi_t$ denotes stochastic decoding randomness. 
The finalized variant is selected according to the total reward in Eq.~\eqref{eq:total_reward}:
\begin{equation}
Z_t
=
\operatorname{Select}
\left(
\{\tilde{Z}_t^{(g)}\}_{g=1}^{G},
R_{\mathrm{total}}
\right).
\end{equation}
Then the cache and queue are updated:
\begin{equation}
\mathcal{C}_{t}
=
\mathcal{C}_{t-1}\cup\{(X_t,Z_t)\},
\end{equation}
\begin{equation}
\mathcal{Q}_{t}
=
\operatorname{UpdateQueue}(\mathcal{Q}_{t-1},X_t).
\end{equation}

The recursively coupled process can be summarized as
\begin{equation}
\begin{aligned}
\pi
\rightarrow
\mathcal{C}_{t-1},\mathcal{Q}_{t-1}
\rightarrow
\mathcal{H}_t
\rightarrow
p_{\theta_t}(Z_t|X_t,\mathbf{U}_t,\mathcal{H}_t)
\\
\quad
\rightarrow
Z_t
\rightarrow
\mathcal{C}_t,\mathcal{Q}_t .
\label{eq:app_recursive_process}
\end{aligned}
\end{equation}
We denote the residual hidden trajectory for the target step by
\begin{equation}
\label{eq:app_gamma}
\Gamma_t
=
(\pi_{1:t},\mathcal{C}_{0:t-1},\mathcal{Q}_{0:t-1},\mathcal{H}_{1:t},\xi_{1:t}).
\end{equation}
For notational simplicity, we omit the step index and write
\begin{equation}
Z\sim p(\cdot|X,\mathcal{O},\Gamma).
\end{equation}

This definition avoids treating released variants themselves as hidden. 
The attacker may observe the released variants, but this does not reveal the ordered trajectory of source-variant pairs used in the cache and prompt. 
Therefore, the attacker must marginalize over possible residual trajectories.

\subsection{Trajectory Leakage Decomposition}

\noindent\textbf{Lemma~\ref{lem:traj} (Trajectory Leakage Decomposition).}
For the obfuscated output $Z\sim p(\cdot|X,\mathcal{O},\Gamma)$,
\begin{align}
I(X;Z|\mathcal{O})=
&\ H(Z|\mathcal{O})
- \mathbb{E}\big[H(Z|X,\mathcal{O},\Gamma)\big] \notag \\
&- I(Z;\Gamma|X,\mathcal{O}).
\end{align}

\nosection{Proof}
By definition,
\begin{equation}
I(X;Z|\mathcal{O})
=
H(Z|\mathcal{O})
-
H(Z|X,\mathcal{O}).
\end{equation}
For the residual trajectory $\Gamma$,
\begin{equation}
I(Z;\Gamma|X,\mathcal{O})
=
H(Z|X,\mathcal{O})
-
\mathbb{E}\big[H(Z|X,\mathcal{O},\Gamma)\big].
\end{equation}
Thus,
\begin{equation}
H(Z|X,\mathcal{O})
=
\mathbb{E}\big[H(Z|X,\mathcal{O},\Gamma)\big]
+
I(Z;\Gamma|X,\mathcal{O}).
\end{equation}
Substituting this equality into the definition of $I(X;Z|\mathcal{O})$ gives the result.
\hfill$\square$

The term $I(Z;\Gamma|X,\mathcal{O})$ measures how much the output distribution changes across residual generation trajectories. 
For a fixed pair $(x,\mathcal{O})$, define the trajectory-mixture distribution as
\begin{equation}
p_{x,\mathcal{O}}(z)
=
\sum_{\gamma}
\pi(\gamma|x,\mathcal{O})
p(z|x,\mathcal{O},\gamma).
\end{equation}
Then
\begin{equation}
\begin{aligned}
I(Z;\Gamma|X=x,\mathcal{O})
=
&\sum_{\gamma}
\pi(\gamma|x,\mathcal{O}) \\
&\cdot
\mathrm{KL}
\Big(
p(\cdot|x,\mathcal{O},\gamma)
\|
p_{x,\mathcal{O}}(\cdot)
\Big).
\end{aligned}
\end{equation}
Therefore, if different residual trajectories induce different output distributions, the attacker faces additional uncertainty from marginalizing over $\Gamma$. 
If the exact processing order, cache states, queue states, source-variant alignments, candidate groups, and sampling path are all revealed or inferable, then they are absorbed into $\mathcal{O}$ and the remaining trajectory term vanishes.

\subsection{Semantic-Mode Entropy}

The following analysis does not assume that the LLM explicitly samples a discrete latent variable. 
Instead, semantic modes are introduced as an analytical partition of the output space. 
For a fixed condition $(x,\mathcal{O},\gamma)$, let
%\begin{equation}
$\{\mathcal{Z}_1,\ldots,\mathcal{Z}_K\}$
%\end{equation}
be a hard or soft partition of possible outputs according to distinguishable semantic directions of obfuscation. 
Let $M\in\{1,\ldots,K\}$ denote the semantic-mode assignment of $Z$.

For each fixed condition $(x,\mathcal{O},\gamma)$, define
\begin{equation}
\alpha_m
=
P(M=m|x,\mathcal{O},\gamma),
\end{equation}
and
\begin{equation}
p_m(\cdot)
=
P(Z=\cdot|M=m,x,\mathcal{O},\gamma).
\end{equation}
Here $\alpha_m$ is the weight of semantic mode $m$, and $p_m$ is the intra-mode output distribution.

\noindent\textbf{Lemma~\ref{lem:mode} (Semantic-Mode Entropy).}
Partition the output space into semantic modes with weights $\alpha_m$ and intra-mode distributions $p_m$. 
Then
\begin{equation}
\begin{split}
\mathbb{E}\big[H(Z|X,\mathcal{O},\Gamma)\big]
&= \mathbb{E}\Big[ H(\alpha) + \sum_m \alpha_m H(p_m) \Big] \\
&\qquad - \epsilon,
\end{split}
\end{equation}
where
\begin{equation}
\epsilon=\mathbb{E}[H(M|Z,X,\mathcal{O},\Gamma)].
\end{equation}

\nosection{Proof}
For fixed $(x,\mathcal{O},\gamma)$, applying the chain rule to $(M,Z)$ gives
{\small
\begin{equation}
\begin{aligned}
H(M,Z|x,\mathcal{O},\gamma)
&=
H(M|x,\mathcal{O},\gamma)
+
H(Z|M,x,\mathcal{O},\gamma),\\
H(M,Z|x,\mathcal{O},\gamma)
&=
H(Z|x,\mathcal{O},\gamma)
+
H(M|Z,x,\mathcal{O},\gamma).
\end{aligned}
\end{equation}}
Since $M$ follows the categorical distribution $\alpha$,
\begin{equation}
H(M|x,\mathcal{O},\gamma)=H(\alpha),
\end{equation}
and
\begin{equation}
H(Z|M,x,\mathcal{O},\gamma)
=
\sum_m \alpha_mH(p_m).
\end{equation}
Combining the two decompositions yields
\begin{equation}
\begin{aligned}
H(Z|x,\mathcal{O},\gamma)
=
&\ H(\alpha)
+
\sum_m \alpha_mH(p_m) \\
&\quad -
H(M|Z,x,\mathcal{O},\gamma).
\end{aligned}
\end{equation}
Taking expectation over $(X,\mathcal{O},\Gamma)$ gives the result.
\hfill$\square$

If the semantic partition is deterministic, then $M$ is determined by $Z$ and $H(M|Z,x,\mathcal{O},\gamma)=0$. 
For soft or overlapping semantic modes, the residual uncertainty is positive and is captured by $\epsilon$. 
Therefore, $\epsilon$ measures the ambiguity of assigning a generated output to a semantic mode.

\subsection{Attack Success Bound}

The rewards do not directly optimize the entropy terms in Lemma~\ref{lem:mode}. 
Instead, they shape the generation policy and affect the leakage decomposition through semantic diversity, intra-mode uncertainty, residual mode ambiguity, and trajectory uncertainty. 
Let $p_{\lambda}$ denote the generation distribution induced by the reward-shaped policy optimized with Eq.~\eqref{eq:total_reward}. 
Under $p_{\lambda}$, the corresponding semantic-mode weights, intra-mode distributions, residual ambiguity, and trajectory divergence are denoted by
$\alpha_m^{\lambda},
p_m^{\lambda},
\epsilon^{\lambda},
I_{\lambda}(Z;\Gamma|X,\mathcal{O})$.

\noindent\textbf{Theorem~\ref{thm:bound} (Attack Success Bound for PGPO).}
For a finite original-word set $\mathcal{X}$ and an approximately uniform prior

\begin{small}
\begin{align}
P_{\mathrm{succ}}
\le
\frac{1}{\log_2|\mathcal{X}|}
\Bigg(
& \epsilon^{\lambda}
- \mathbb{E}_{\lambda}\!\left[
H(\alpha^{\lambda})
+\sum_m \alpha_m^{\lambda}H(p_m^{\lambda})
\right] \notag \\
&+ H_{\lambda}(Z|\mathcal{O})
- I_{\lambda}(Z;\Gamma|X,\mathcal{O})
+ 1
\Bigg).
\end{align}
\end{small}

\nosection{Proof}
Substituting Lemma~\ref{lem:mode} into Lemma~\ref{lem:traj} gives
\begin{equation}
\label{eq:app_leakage_decomp}
\begin{aligned}
I_{\lambda}(X;Z|\mathcal{O})
=
&
H_{\lambda}(Z|\mathcal{O})
+
\epsilon^{\lambda}
-
I_{\lambda}(Z;\Gamma|X,\mathcal{O})\\
&-
\mathbb{E}_{\lambda}
\left[
H(\alpha^{\lambda})
+
\sum_m
\alpha_m^{\lambda}H(p_m^{\lambda})
\right].
\end{aligned}
\end{equation}
Let $\widehat{X}(Z,\mathcal{O})$ be the attacker's estimator of the original input $X$, and let
\begin{equation}
P_{\mathrm{succ}}
=
\Pr[\widehat{X}(Z,\mathcal{O})=X].
\end{equation}
For a finite candidate space $\mathcal{X}$ and an approximately uniform prior over $X$ after conditioning on $\mathcal{O}$, Fano's inequality~\citep{fano1961transmission,nalewajski2011elements} gives
\begin{equation}
P_{\mathrm{succ}}
\le
\frac{
I_{\lambda}(X;Z|\mathcal{O})+1
}
{\log_2|\mathcal{X}|}.
\end{equation}
Substituting Eq.~\eqref{eq:app_leakage_decomp} into the above inequality gives the result.
\hfill$\square$

The bound shows that PGPO can reduce attack success by increasing semantic-mode diversity, preserving valid intra-mode uncertainty, controlling residual mode ambiguity, and forcing the attacker to marginalize over residual generation trajectories. 
When $H_{\lambda}(Z|\mathcal{O})$ is controlled by the output format and category constraints, increasing the negative terms in the bound tightens the attack success upper bound.

\subsection{Reward Interpretation}

The reward-induced bound explains the roles of the three rewards. 
The individual dissimilarity reward in Eq.~\eqref{eq:reward_dis} encourages variants to move away from the original word along diverse semantic directions. 
This tends to increase the empirical coverage of semantic modes and hence the mode-weight entropy $H(\alpha^{\lambda})$.

The structural consistency reward in Eq.~\eqref{eq:reward_str} preserves local relational order among history samples. 
By making semantic modes more consistently assignable to relationally meaningful regions, it helps control the residual ambiguity $\epsilon^{\lambda}$, which appears as a positive term in the attack success bound.

The semantic constraint reward in Eq.~\eqref{eq:reward_con} anchors generated variants to category-valid semantic manifolds. 
It prevents diversity from drifting into invalid or out-of-category regions, thereby preserving meaningful intra-mode uncertainty 
$\sum_m \alpha_m^{\lambda}H(p_m^{\lambda})$ rather than producing degenerate noise.

Overall, the reward effects can be summarized as
\begin{equation}
\begin{aligned}
R_{\mathrm{dis}} &\Rightarrow H(\alpha^{\lambda}), \\
R_{\mathrm{str}} &\Rightarrow \epsilon^{\lambda}, \\
R_{\mathrm{con}} &\Rightarrow \sum_m \alpha_m^{\lambda}H(p_m^{\lambda}).
\end{aligned}
\label{eq:regularization_mapping}
\end{equation}
These arrows indicate mechanism-level influence rather than mathematical equivalence.

\subsection{Representation-level Leakage}

Finally, we connect text-level obfuscation to the final item representation used by recommender models. 
Let
\begin{equation}
\mathbf{h}_{\mathrm{item}}
=
g(e(Z),S),
\end{equation}
where $e(Z)$ is the text embedding of the obfuscated variant and $S$ denotes collaborative signals. 
Let $C$ denote category or contextual information used in the representation process. 
Since $\mathbf{h}_{\mathrm{item}}$ is obtained from $(e(Z),S)$ through the representation function $g$, the data processing inequality gives Eq.~\eqref{eq:rep_dpi}.

By the chain rule,
\begin{equation}
\label{eq:app_chain_rep}
\begin{aligned}
I(X;e(Z),S|C,\mathcal{O})
=
&\ I(X;e(Z)|C,\mathcal{O}) \\
&+
I(X;S|e(Z),C,\mathcal{O}).
\end{aligned}
\end{equation}
If collaborative signals contain limited residual information about $X$, we assume
\begin{equation}
\label{eq:app_collab_residual}
I(X;S|e(Z),C,\mathcal{O})
\le
\eta.
\end{equation}
If the downstream representation introduces at most $\delta$ additional approximation slack and the text branch contributes with effective coefficient $\rho_\alpha$, then Eq.~\eqref{eq:rep_bound} follows.

Therefore, reducing text-level leakage also reduces the information available through the final item representation, except for residual collaborative information and model approximation slack.

\section{Experimental Settings}
\label{app:experimental_settings}

\subsection{Dataset Preprocessing}
\label{app:dataset_preprocessing}

\nosection{MovieLens-1M}
We keep ratings $\ge 3$ as positive feedback and retain only the latest interaction for duplicate user-item pairs. Each item keeps one anchor interaction, while the remaining interactions are randomly kept with probability 0.4. Users with fewer than 3 interactions are fully assigned to the training set.

\nosection{Amazon-Book}
Interactions from January to March 2018 are first filtered to retain users and items with at least 3 interactions. We then randomly sample 10,000 items and keep their associated interactions. After sampling, users and items with fewer than 3 interactions are removed again. Ratings $\ge 4$ are treated as positive feedback.

\subsection{Item-side Attribute Construction}
\label{app:attribute_construction}

We use Qwen2.5-14B to generate fine-grained, multi-category textual attributes from item titles.
For each item $i$, the LLM yields an attribute set $\mathcal{A}_i$, where each attribute word is assigned to a category $c\in\mathcal{C}$.
Attributes are organized per category as $\mathrm{ItemEdges}[i]=\{c:\mathcal{W}_{i,c}\}$, and the initial vocabulary is $\mathcal{W}_{\mathrm{init}}=\bigcup_i\mathcal{A}_i$.

\subsubsection{Prompt Templates}
\label{app:prompt_templates}

\noindent
\begin{minipage}{\columnwidth}
\begin{lstlisting}[
basicstyle=\ttfamily\small,
breaklines=true
]
# MovieLens-1M Movie Prompt
You are a movie database curator.

Introduce the movie titled "{item_name}"
in valid JSON formatCritical Acclaim",
"Production Quality", "Soundtrack"

For each key, output 3-5 concise,
distinctive English keywords.
Each keyword should be 2-5 words long
and discriminative.

Output ONLY the JSON object.
\end{lstlisting}
\end{minipage}

\medskip

\noindent
\begin{minipage}{\columnwidth}
\begin{lstlisting}[
basicstyle=\ttfamily\small,
breaklines=true
]
# Amazon-Book Prompt
You are a book metadata curator.

Introduce the book titled "{item_name}"
in valid JSON format with exactly these 9 keys:
"Genre", "Author", "PublicationContext",
"CharacterOrFocus", "PlotAndTheme",
"ToneAndStyle", "AudienceAndUseCase",
"DistinctiveElements", "RelatedTopics"

For each key, output 3-5 concise,
distinctive English keywords.

Output ONLY the JSON object.
\end{lstlisting}
\end{minipage}

\subsubsection{Category and Keyword Pruning}
\label{app:category_keyword_pruning}

After generating attributes, we prune redundant categories and retain discriminative keywords by:
(1) computing IDF~\citep{salton1988term} for each keyword,
(2) scoring each category by average keyword IDF and pairwise embedding similarity,
(3) iteratively removing categories with high redundancy and low score, and
(4) keeping the top-$\tau$ keywords per retained category for each item.
The resulting retained categories and keyword counts are reported in Table~\ref{tab:retained_keywords_three_col}.

\begin{table}[t]
\centering
\footnotesize
\renewcommand{\arraystretch}{1.0}

\begin{tabular*}{\columnwidth}{
@{\hspace{4pt}\extracolsep{\fill}}
c c c
@{\hspace{4pt}}
}
\toprule
\textbf{Dataset}
& \textbf{Category}
& \textbf{\#Keywords} \\
\midrule

\multirow{5}{*}{MovieLens-1M}
& PlotAndTheme       & 6,384  \\
& Actors             & 5,122  \\
& Soundtrack         & 4,039  \\
& Production Quality & 2,893  \\
& \textbf{Total}     & \textbf{18,398} \\
\midrule

\multirow{5}{*}{Amazon-Book}
& PlotAndTheme        & 7,637  \\
& Author              & 6,242  \\
& PublicationContext  & 4,513  \\
& ToneAndStyle        & 3,725  \\
& \textbf{Total}      & \textbf{21,678} \\
\bottomrule
\end{tabular*}

\caption{Retained categories and keyword counts after category pruning.
The total differs from the sum of the individual categories because some
keywords appear in multiple categories.}
\label{tab:retained_keywords_three_col}
\end{table}

\subsection{Baselines}
\label{app:baselines}

We compare our method with the following baselines:
\begin{itemize}[noitemsep, topsep=0pt]
    \item \textbf{None:} Item embeddings constructed without using side information, serving as the lower bound.
    \item \textbf{Description:} Raw side information, serving as the utility upper bound.
    \item \textbf{InferDPT:} A differentially private text perturbation method that injects Laplacian noise during text encoding~\citep{tong2025inferdpt}, used directly without LLM processing.
    \item \textbf{Text2Text:} A sequence-to-sequence text obfuscation method that perturbs embeddings under local differential privacy and decodes them back~\citep{feyisetan2020privacy,qu2021natural}.
    \item \textbf{SanText+:} An improved text obfuscation method that optimizes substitution word selection to enhance obfuscation quality~\citep{yue2021differential}.
    \item \textbf{TextObfuscator (TOF):} A rule-based text obfuscation tool that replaces sensitive words for privacy protection~\citep{zhou2023textobfuscator}.
\end{itemize}

\subsection{Privacy Attack Settings}
\label{app:attack_model}
\nosection{Embedding Inversion Attack}
We adopt the Vec2Text embedding inversion attack~\citep{morris2023text} to recover sensitive category labels from item embeddings. The attack training data does not overlap with the items used in the main recommendation experiments. For MovieLens-1M, we sample 80,000 movie tags from MovieLens-25M\footnote{\url{https://grouplens.org/datasets/movielens/25m/}} and  MovieLens-32M\footnote{\url{https://grouplens.org/datasets/movielens/32m/}}. For Amazon-Book, we sample 100,000 disjoint items. Embeddings are generated using the same pipeline as in the main experiments.

Since different obfuscation methods induce different degrees of semantic displacement, directly comparing attack success rates can be misleading. For fair evaluation, we tune the hyperparameters of each method, such as dissimilarity reward and DP noise, so that the average cosine similarity between original and obfuscated embeddings is aligned. All inversion attacks are evaluated under this aligned CosSim setting.

\nosection{Maximum-likelihood Inversion Attack}
We consider a strong white-box attacker with access to the optimized
conversion model, original vocabulary, static semantic graph, retrieval
function $\phi$, and generation prompt. The attacker observes the released
variant $z$, but not the cached history $\mathcal{H}$ or the realized
generation trajectory $\Gamma$.

For each observed variant, we construct a candidate set $\mathcal{C}$
containing the ground-truth source word and 50 sampled negatives. For each
candidate $x_i\in\mathcal{C}$, the attacker replays the generation prompt
with $\mathcal{H}=\varnothing$ and computes the teacher-forced likelihood
of $z$.

\subsection{Feature Processing Pipeline}
\label{app:feature_processing}

All original and obfuscated texts are encoded into 768-dimensional embeddings using BERT-base-uncased. For traditional recommendation models, PCA is applied to reduce the dimension to 128. For multimodal and LLM-based models, the full 768-dimensional embeddings are retained and integrated through learnable projection layers.

\subsection{Training Configuration}
\label{app:training_configuration}

For both datasets, the prototype word count is set to $5000$, the cache size is set to $2000$, the reward weights are
$(\lambda_{\text{dis}}, \lambda_{\text{str}}, \lambda_{\text{con}})=(1,2,1)$, and the warm-up length is
\begin{equation}
T_{\text{warmup}}=2\left\lceil \frac{|C_{\text{Prototype}}|}{\mathrm{batch\_size}} \right\rceil.
\end{equation}

We adopt a phased reward schedule. During the first
$\left\lceil |C_{\text{Prototype}}|/\mathrm{batch\_size} \right\rceil$
steps, only the semantic constraint and individual dissimilarity rewards are optimized by setting $\lambda_{\text{str}}=0$. 
After sufficient high-quality anchors are cached, the structural consistency reward is restored to its default weight.

All ablation and parameter sensitivity experiments are conducted on MovieLens-1M. 
For ablation studies, we use the same hyperparameter settings as the MovieLens-1M main experiment unless otherwise specified. 
For parameter sensitivity analysis, we vary one factor at a time while keeping the others fixed. 
In cache size sensitivity, we set
$(\lambda_{\text{dis}}, \lambda_{\text{str}}, \lambda_{\text{con}})=(1,1,1)$
and vary the cache size in $\{0,1000,2000,3000,4000,5000\}$. 
In prototype word count sensitivity, we use
$(\lambda_{\text{dis}}, \lambda_{\text{str}}, \lambda_{\text{con}})=(1,2,1)$
and vary the prototype word count in $\{1250,2500,5000,7500,10000\}$. 
In reward weight sensitivity, we fix the cache size to $2000$ and start from
$(\lambda_{\text{dis}}, \lambda_{\text{str}}, \lambda_{\text{con}})=(1,1,1)$. 
We then vary $\lambda_{\text{dis}}$, $\lambda_{\text{str}}$, and $\lambda_{\text{con}}$ individually, with each weight taking values from $\{1,2,3\}$.

The batch size is set to $2$, and all PGPO training and generation experiments are conducted on a single NVIDIA GeForce RTX 4090 GPU with 24GB memory.

\subsection{Evaluation Metrics}
\label{app:evaluation_metrics}
We evaluate our method from three perspectives: obfuscated variants quality, recommendation performance, and embedding inversion defense

\subsubsection{Obfuscated Variants Quality}
We evaluate generation quality at both the word and context levels.

For word-level evaluation, we encode original words and generated variants with BERT-base-uncased embeddings. We report Distinct-1~\citep{li2016diversity} to measure token-level diversity, and use Spearman correlation to evaluate whether the local semantic structure is preserved after transformation. In addition, we define \textbf{Valid}, \textbf{Spearman} and \textbf{LR} for our task.

\textbf{Valid} measures the proportion of generated variants that are not identical to their original words:
\begin{equation}
\mathrm{Valid}
=
1 -
\frac{1}{N}
\sum_{i=1}^{N}
\mathbb{I}(v_i = o_i),
\end{equation}
where $o_i$ and $v_i$ denote the original word and its generated variant, respectively.

\textbf{Spearman} measures whether the local semantic structure is preserved after transformation. For each original word, we retrieve its top-$k$ nearest neighbors within the same category in the original BERT embedding space, and compare the similarity rankings before and after transformation using Spearman rank correlation~\citep{spearman1961proof}. We average the correlation over all valid samples:
\begin{equation}
\mathrm{Spearman}
=
\frac{1}{N}
\sum_{i=1}^{N}
\rho_i,
\end{equation}
where $\rho_i$ denotes the rank correlation for the $i$-th sample. A higher value indicates better preservation of the local neighborhood structure.

\textbf{LR} measures the average length ratio between generated variants and original words:
\begin{equation}
\mathrm{LR}
=
\frac{1}{N}
\sum_{i=1}^{N}
\frac{\mathrm{len}(v_i)}
{\mathrm{len}(o_i)}.
\end{equation}

For context-level evaluation, we replace the original words in downstream contexts with their generated variants and evaluate the resulting texts. We report Distinct-2~\citep{li2016diversity} to measure contextual diversity. We use Qwen3-4B~\citep{yang2025qwen3} to compute KL divergence~\citep{nalewajski2011elements} and NLL~\citep{bengio2003neural}, where KL measures the average divergence between token probability distributions of original and transformed contexts, and NLL (from LM cross-entropy loss) evaluates the fluency of the transformed contexts.

\subsubsection{Recommendation Performance}
\label{app:recommendation_metrics}
For traditional recommendation models, including collaborative filtering and multimodal recommendation methods, we follow a chronological 8:1:1 data split and report HR@10 and NDCG@10. For SASRec, the latest 20 interacted items are used as the input sequence. For LLM-based recommendation models, we adopt the dataset splits and evaluation metrics from their original papers: TALLRec~\citep{bao2023tallrec} is evaluated with AUC, while LLaRA~\citep{liao2024llara} is evaluated with HR@1. For all experiments, we used the best-performing model on the validation set for testing. All reported results are averaged over 5 random seeds.

\subsubsection{Embedding Inversion Defense}
To evaluate embedding inversion defense, we formulate the problem as multi-label attribute prediction. 
Given the predicted attribute set $P_i$ and the ground-truth attribute set $T_i$ for each sample, we evaluate attack and defense effectiveness with standard multi-label metrics, including EM, Precision, Recall, F1, and Jaccard. 
EM measures whether the predicted attribute set exactly matches the ground-truth set. 
Precision, Recall, F1, and Jaccard measure prediction correctness, coverage, their harmonic mean, and set-level similarity, respectively.

\subsubsection{White-box Re-identification}

We evaluate vocabulary-alignment resistance using the
Replay-MLE-NoHistory white-box attack. For each released variant
$z_n$, we construct a candidate set $\mathcal{C}_n$ of size $M$,
containing the ground-truth source word $x_n^{*}$ and sampled negative
words. The attacker reuses the optimized checkpoint and prompt
template, while setting the unavailable history to
$\mathcal{H}=\varnothing$. For each candidate
$x_i\in\mathcal{C}_n$, its teacher-forced likelihood score is
\begin{equation}
s_{n,i}
=
\log p_{\theta}
\left(
z_n
\mid
\mathrm{Prompt}(x_i,c_i),
\mathcal{H}=\varnothing
\right),
\end{equation}
where $c_i$ denotes the category associated with $x_i$. The scores are converted into an attack-induced posterior distribution:
\begin{equation}
q_{n,i}
=
q_{\mathrm{att}}
\left(
x_i\mid z_n,\mathcal{O}
\right)
=
\frac{\exp(s_{n,i}/T)}
{\sum_{x_j\in\mathcal{C}_n}\exp(s_{n,j}/T)},
\end{equation}
where $T$ is the posterior temperature. Candidates are ranked in
descending order of $s_{n,i}$, and $r_n$ denotes the rank of the
ground-truth source word $x_n^{*}$.

\textbf{ReID@1} measures the proportion of variants for which the
ground-truth source word is ranked first:

\begin{equation}
\mathrm{ReID@1}
=
\frac{1}{N}
\sum_{n=1}^{N}
\mathbb{I}(r_n=1).
\end{equation}

\textbf{MRR} measures the average reciprocal rank of the ground-truth
source word:

\begin{equation}
\mathrm{MRR}
=
\frac{1}{N}
\sum_{n=1}^{N}
\frac{1}{r_n}.
\end{equation}

Lower ReID@1 and MRR indicate stronger resistance to recovering the
source--variant correspondence.

\textbf{Posterior entropy} measures the uncertainty of the
attack-induced posterior. For each released variant, it is defined as

\begin{equation}
H_{\mathrm{att}}^{(n)}
=
-
\sum_{x_i\in\mathcal{C}_n}
q_{n,i}\log_2 q_{n,i}.
\end{equation}

We report its average over all evaluated variants:

\begin{equation}
H_{\mathrm{att}}
=
\frac{1}{N}
\sum_{n=1}^{N}
H_{\mathrm{att}}^{(n)}.
\end{equation}

This quantity serves as an operational proxy for
$H(X\mid Z,\mathcal{O})$, with a higher value indicating greater
attacker uncertainty.

\textbf{MI proxy} measures the information revealed by the released
variant relative to the candidate prior:

\begin{equation}
I_{\mathrm{att}}
=
H(X\mid\mathcal{O})
-
H_{\mathrm{att}}.
\end{equation}

Under an approximately uniform prior over a candidate set of size $M$,
we have $H(X\mid\mathcal{O})=\log_2 M$, and therefore

\begin{equation}
I_{\mathrm{att}}
=
\log_2 M
-
H_{\mathrm{att}}.
\end{equation}

Thus, higher $H_{\mathrm{att}}$ and lower $I_{\mathrm{att}}$ indicate
stronger source--variant unlinkability. Since exhaustive scoring over
the full vocabulary is computationally expensive, all metrics are
estimated under a fixed-size sampled candidate protocol.

\subsection{Artifact Use}
We use publicly available datasets, pretrained models, recommendation models, attack tools, and baseline implementations only for academic research and evaluation, following their corresponding licenses, terms of use, and intended research purposes. 
The derived side-information attributes and obfuscated variants are used solely for privacy-preserving recommendation experiments and are not intended for commercial deployment, user profiling, or redistribution outside research contexts. 
Our released code includes license information and reproduction instructions under the access conditions of the original datasets and models.

\section{Additional Analysis}
\label{app:additional_analysis}
\begin{figure}[t]
  \centering
  \begin{minipage}[t]{0.48\linewidth}  
    \centering
    \includegraphics[width=\linewidth]{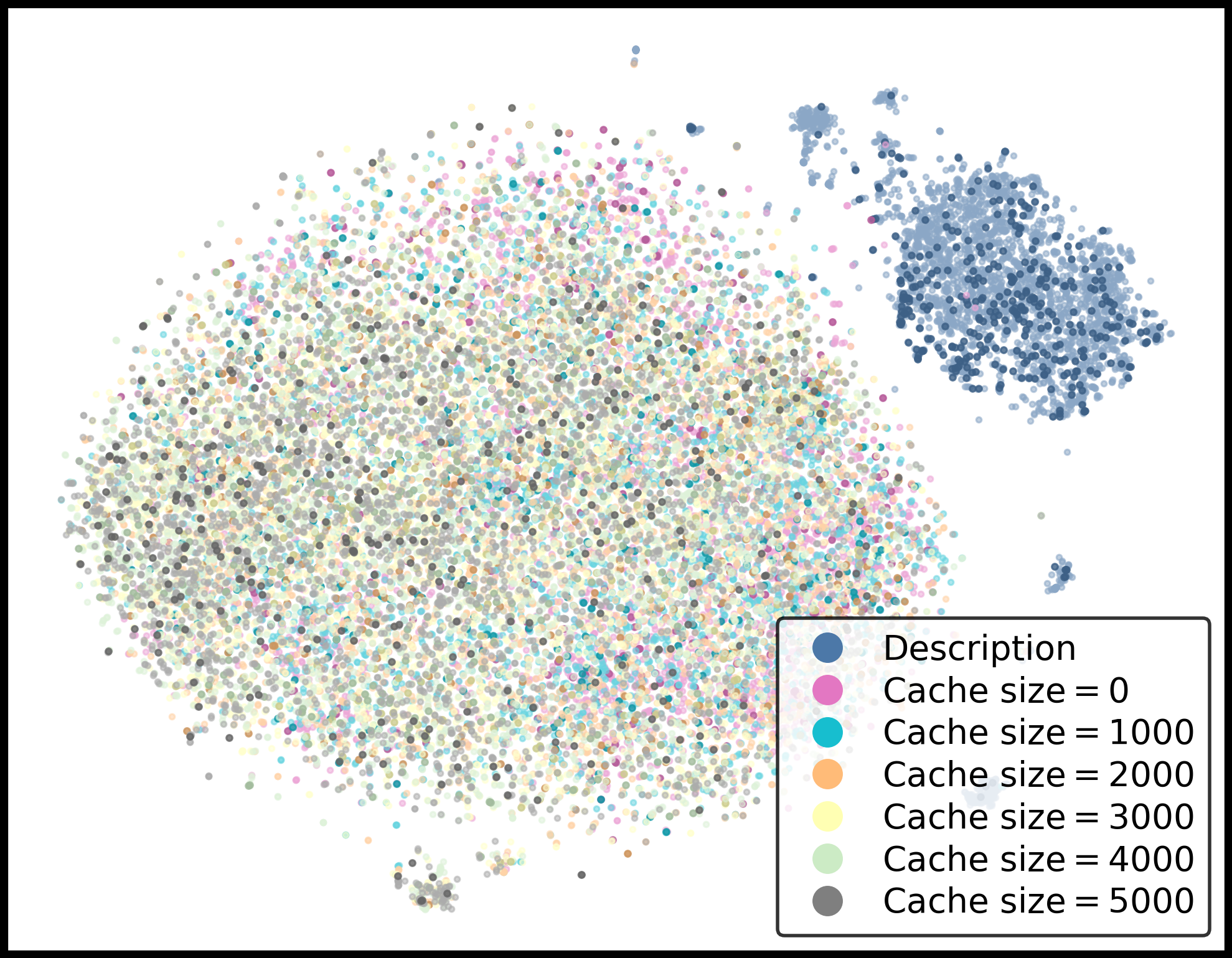}
    \caption{T-SNE of embeddings across cache sizes.}
    \label{fig:cache_tsne}
  \end{minipage}%
  \hfill
\begin{minipage}[t]{0.48\linewidth}  
    \centering
    \includegraphics[width=\linewidth]{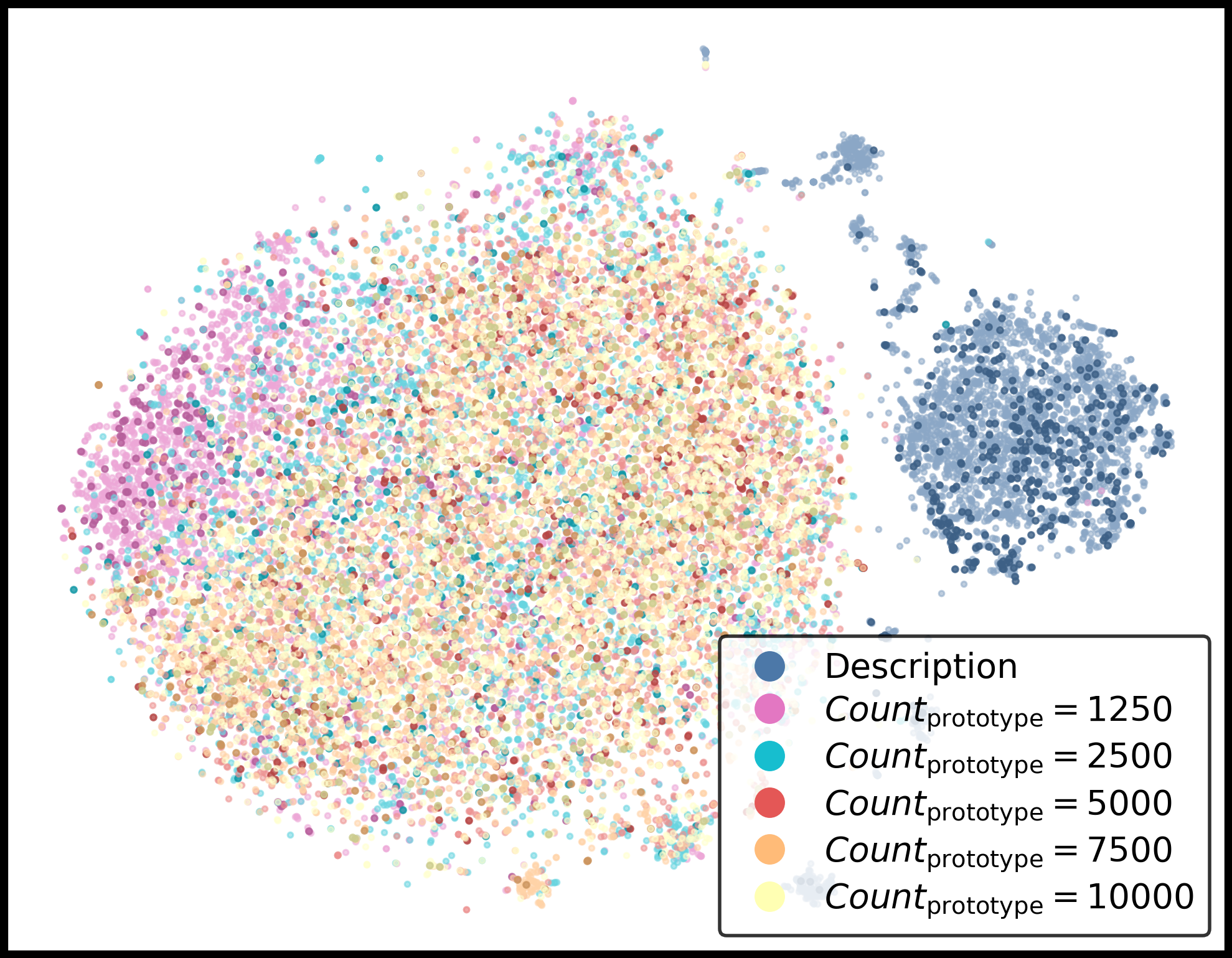}
    \caption{T-SNE of embeddings across prototype word counts.}
    \label{fig:core_word_tsne}
  \end{minipage}%
\end{figure}

\begin{table*}[t]
\centering
\footnotesize
\renewcommand{\arraystretch}{0.9}

%\begin{tabular*}{\textwidth}{@{\extracolsep{\fill}}ccccccccc}
\begin{tabular*}{\textwidth}{
@{\hspace{8pt}\extracolsep{\fill}}
ccccccccc
@{\hspace{8pt}}
}
\toprule
\textbf{Cache Size}
& \textbf{Valid}$^{\dagger}$
& \textbf{CosSim}$^{\dagger}$
& \textbf{KL}$^{\ddagger}$
& \textbf{Spearman}$^{\dagger}$
& \textbf{NLL}$^{\ddagger}$
& \textbf{D-2}$^{\ddagger}$
& \textbf{D-1}$^{\dagger}$
& \textbf{LR}$^{\dagger}$ \\
\midrule
0    & 0.9982 & 0.6333 & 8.1727 & 0.0070 & 4.3177 & 0.3828 & 0.1430 & 0.9021 \\
1000 & 0.9990 & 0.6290 & 7.9774 & 0.1771 & 4.3254 & 0.3996 & 0.1668 & 0.8787 \\
2000 & 0.9991 & 0.6241 & 7.9014 & 0.1728 & 5.8615 & 0.4112 & 0.1783 & 0.8706 \\
3000 & 0.9990 & 0.6178 & 7.7505 & 0.1672 & 6.7148 & 0.4327 & 0.2079 & 0.8439 \\
4000 & 0.9994 & 0.6155 & 7.7191 & 0.1655 & 7.2066 & 0.4469 & 0.2278 & 0.8571 \\
5000 & 0.9996 & 0.6120 & 7.7077 & 0.1653 & 8.0208 & 0.4456 & 0.2258 & 0.8619 \\
\bottomrule
\end{tabular*}

\vspace{2pt}
\parbox{\textwidth}{%
\footnotesize
$^{\dagger}$ Word-level metrics;
$^{\ddagger}$ Item-description-level metrics.
}

\caption{Parameter sensitivity analysis on cache size.}
\label{tab:cache_psa}
\end{table*}

\begin{table*}[t]
\centering
\footnotesize
\renewcommand{\arraystretch}{0.9}

%\begin{tabular*}{\textwidth}{@{\extracolsep{\fill}}ccccccccc}
\begin{tabular*}{\textwidth}{
@{\hspace{8pt}\extracolsep{\fill}}
ccccccccc
@{\hspace{8pt}}
}
\toprule
\textbf{Prototype Count}
& \textbf{Valid}$^{\dagger}$
& \textbf{CosSim}$^{\dagger}$
& \textbf{KL}$^{\ddagger}$
& \textbf{Spearman}$^{\dagger}$
& \textbf{NLL}$^{\ddagger}$
& \textbf{D-2}$^{\ddagger}$
& \textbf{D-1}$^{\dagger}$
& \textbf{LR}$^{\dagger}$ \\
\midrule
1250  & 0.9951 & 0.6309 & 7.3785 & 0.2352 & 8.3621 & 0.4705 & 0.3270 & 0.7863 \\
2500  & 0.9970 & 0.6442 & 7.5392 & 0.2353 & 6.9372 & 0.4261 & 0.2240 & 0.8422 \\
5000  & 0.9986 & 0.6293 & 7.9264 & 0.1981 & 5.7050 & 0.4257 & 0.1960 & 0.8758 \\
7500  & 0.9976 & 0.6228 & 7.7439 & 0.1650 & 6.8084 & 0.4333 & 0.2117 & 0.8518 \\
10000 & 0.9986 & 0.6209 & 7.7050 & 0.1308 & 7.2492 & 0.4576 & 0.2323 & 0.8692 \\
\bottomrule
\end{tabular*}

\vspace{2pt}
\parbox{\textwidth}{%
\footnotesize
$^{\dagger}$ Word-level metrics;
$^{\ddagger}$ Item-description-level metrics.
}

\caption{Parameter sensitivity analysis on prototype count.}
\label{tab:coreword_psa}
\end{table*}

\begin{table*}[t]
\centering
\footnotesize
\renewcommand{\arraystretch}{0.9}
\begin{tabular*}{\textwidth}{
@{\hspace{8pt}\extracolsep{\fill}}
ccccccccc
@{\hspace{8pt}}
}
%\begin{tabular*}{\textwidth}{@{\extracolsep{\fill}}ccccccccc}
\toprule$\boldsymbol{(\lambda_{\mathrm{dis}},\lambda_{\mathrm{str}},\lambda_{\mathrm{con}})}$
& \textbf{Valid}$^{\dagger}$
& \textbf{CosSim}$^{\dagger}$
& \textbf{KL}$^{\ddagger}$
& \textbf{Spearman}$^{\dagger}$
& \textbf{NLL}$^{\ddagger}$
& \textbf{D-2}$^{\ddagger}$
& \textbf{D-1}$^{\dagger}$
& \textbf{LR}$^{\dagger}$ \\
\midrule
$(1,1,1)$ & 0.9991 & 0.6241 & 7.9014 & 0.1728 & 5.8615 & 0.4112 & 0.1783 & 0.8706 \\
$(2,1,1)$ & 0.9995 & 0.5740 & 7.8329 & 0.1517 & 7.3188 & 0.4650 & 0.2788 & 0.8191 \\
$(3,1,1)$ & 0.9986 & 0.5036 & 8.3222 & 0.1539 & 7.9570 & 0.5864 & 0.6248 & 0.7775 \\
$(1,2,1)$ & 0.9986 & 0.6293 & 7.9264 & 0.1981 & 5.7050 & 0.4257 & 0.1960 & 0.8758 \\
$(1,3,1)$ & 0.9976 & 0.6324 & 7.6278 & 0.2118 & 6.9764 & 0.4309 & 0.2107 & 0.8537 \\
$(1,1,2)$ & 0.9945 & 0.6624 & 7.6043 & 0.1579 & 6.2802 & 0.4120 & 0.1855 & 0.8589 \\
$(1,1,3)$ & 0.9797 & 0.7001 & 7.4114 & 0.1521 & 6.1095 & 0.4171 & 0.1913 & 0.8755 \\
\bottomrule
\end{tabular*}

\vspace{2pt}
\parbox{\textwidth}{%
\footnotesize
$^{\dagger}$ Word-level metrics;
$^{\ddagger}$ Item-description-level metrics.
}

\caption{Parameter sensitivity analysis on reward weight configuration.}
\label{tab:weight_psa}
\end{table*}

\subsection{Cache Size} 
Fig.~\ref{fig:cache_tsne} shows that the global embedding distribution remains largely stable across cache sizes, while individual samples shift locally. 
As shown in Table~\ref{tab:cache_psa} and Fig.~\ref{fig:psa_cache_size}, removing the cache severely weakens local structure preservation and leads to poor recommendation performance. 
A moderate cache substantially improves Spearman correlation and recommendation accuracy while reducing attack effectiveness, indicating that cached history samples provide useful contextual anchors for preserving local topology and promoting sufficient semantic shift. 
However, further increasing the cache size slightly decreases structural preservation and embedding similarity, while increasing NLL. 
This suggests that an overly large cache may bias generation toward excessively low-similarity variants, leading to diminishing returns for the overall privacy--utility trade-off.

\subsection{Prototype Word Count} 
\begin{figure}[t]
\centering
\begin{subfigure}[t]{0.48\linewidth}
    \centering
    \includegraphics[width=\linewidth]{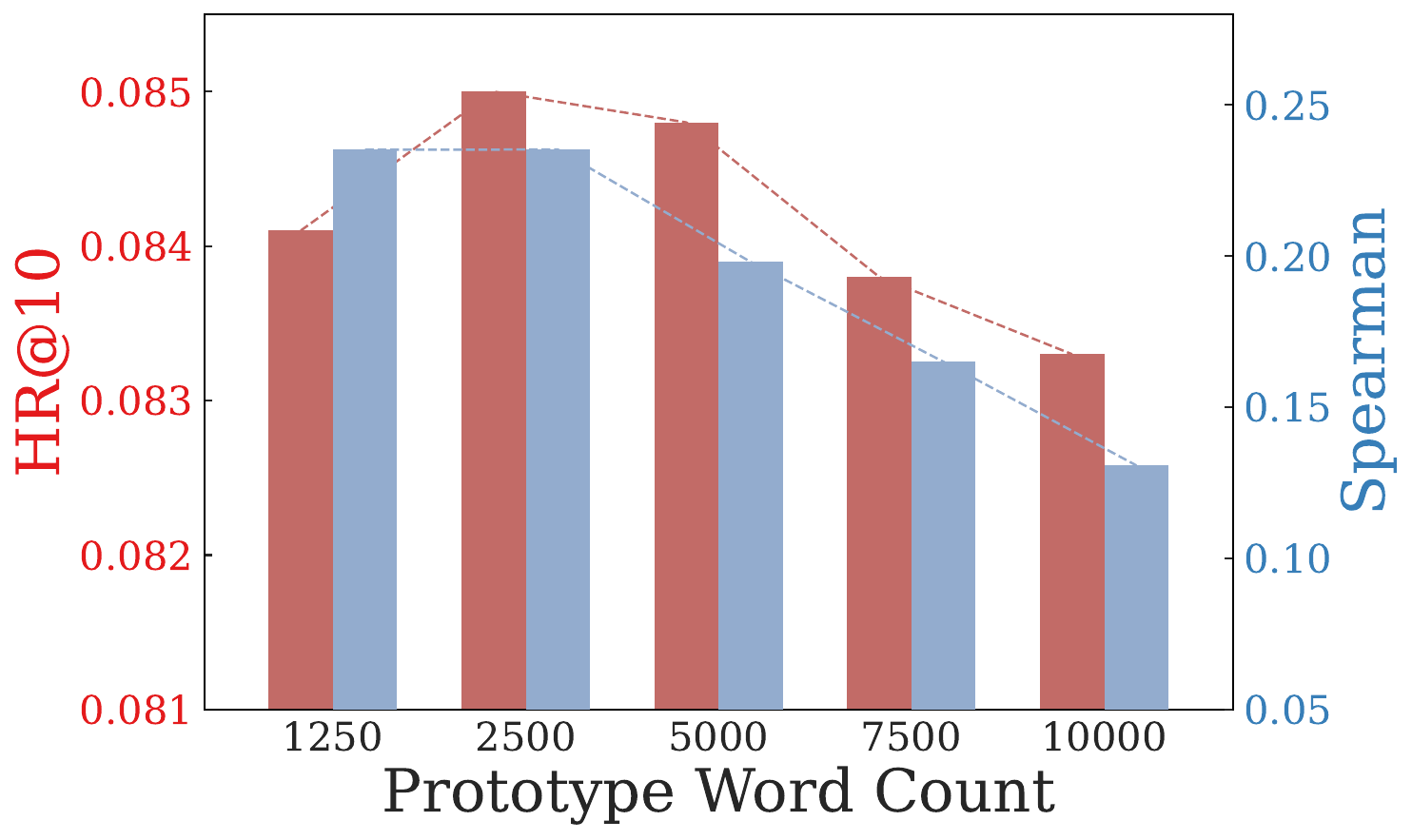}
    \caption{Recommendation utility.}
    \label{fig:psa_sub1}
\end{subfigure}
\hfill
\begin{subfigure}[t]{0.48\linewidth}
    \centering
    \includegraphics[width=\linewidth]{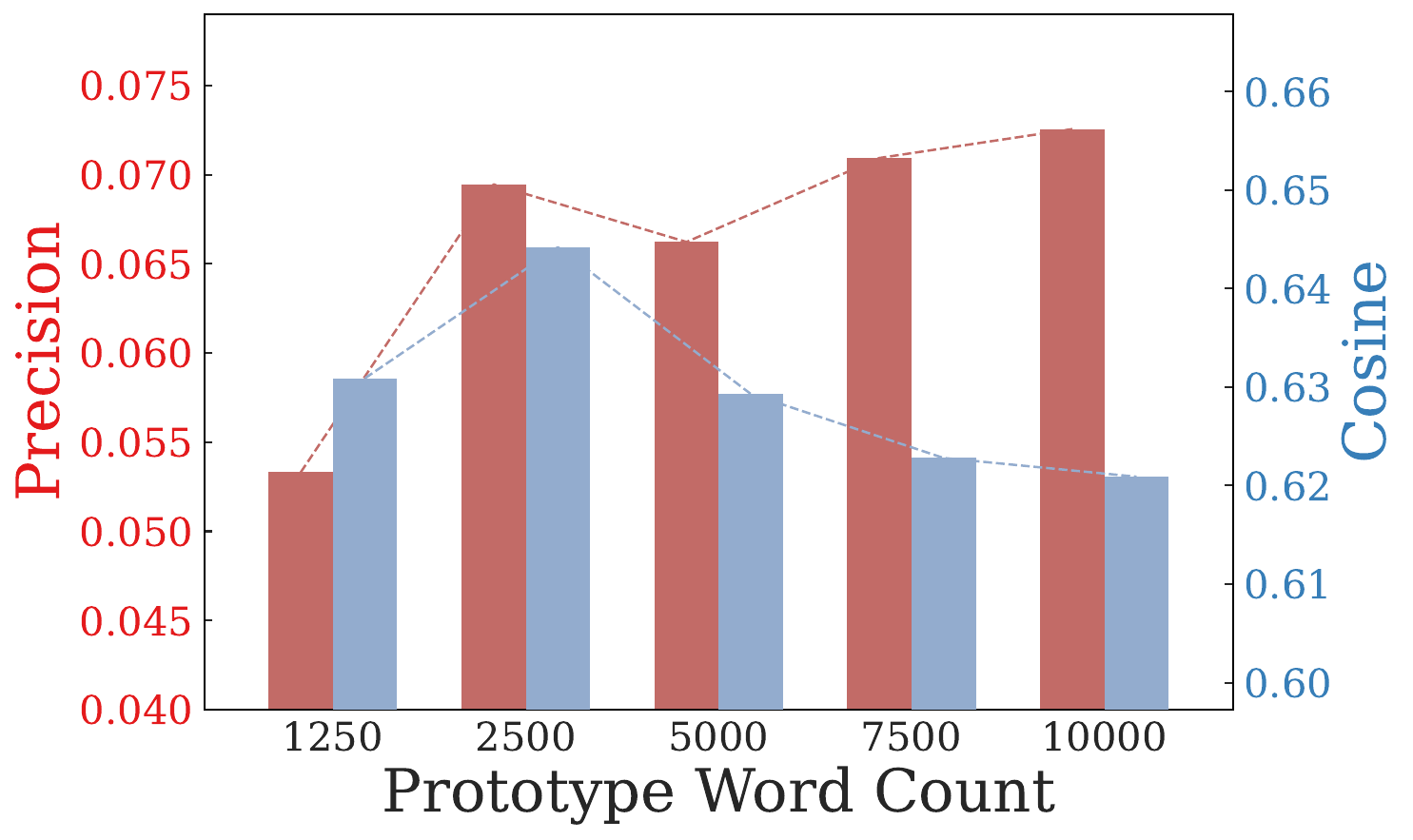}
    \caption{Inversion defence.}
    \label{fig:psa_sub2}
\end{subfigure}
\caption{Prototype Word Count Sensitivity.}
\label{fig:psa_core_word}
\end{figure}

Fig.~\ref{fig:core_word_tsne} shows that the overall embedding layout remains relatively stable across different prototype counts, although individual samples exhibit local shifts. 
As reported in Table~\ref{tab:coreword_psa} and Fig.~\ref{fig:psa_core_word}, a small set of prototypes is already sufficient to establish effective structural anchors, preserving local semantic relations while producing variants with clear semantic displacement. 
Increasing the prototype count moderately improves fluency and slightly benefits recommendation performance. 
However, using too many prototypes makes the structure-learning process harder and encourages stronger semantic deviation, which lowers Spearman correlation and may slightly harm downstream utility. 
These results indicate that prototype selection should emphasize high-influence anchors rather than exhaustive coverage.

\begin{figure*}[ht]
\centering
\begin{subfigure}{0.25\linewidth}
    \centering
    \includegraphics[width=\linewidth]{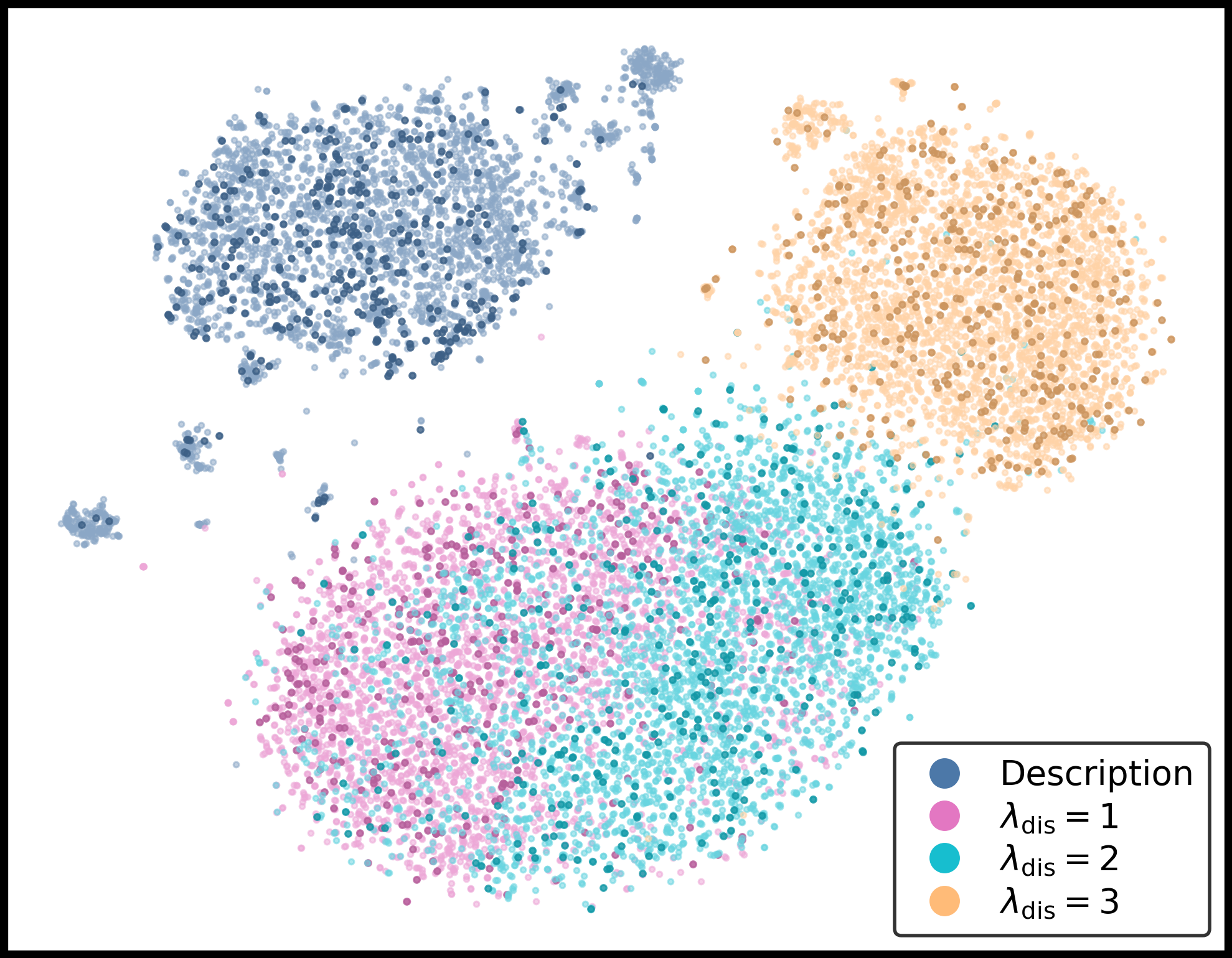}
    \caption{Varying $R_{\mathrm{dis}}$ weight.}
    \label{fig:sub_add_1}
\end{subfigure}
\hspace{0.05\linewidth}% 
\begin{subfigure}{0.25\linewidth}
    \centering
    \includegraphics[width=\linewidth]{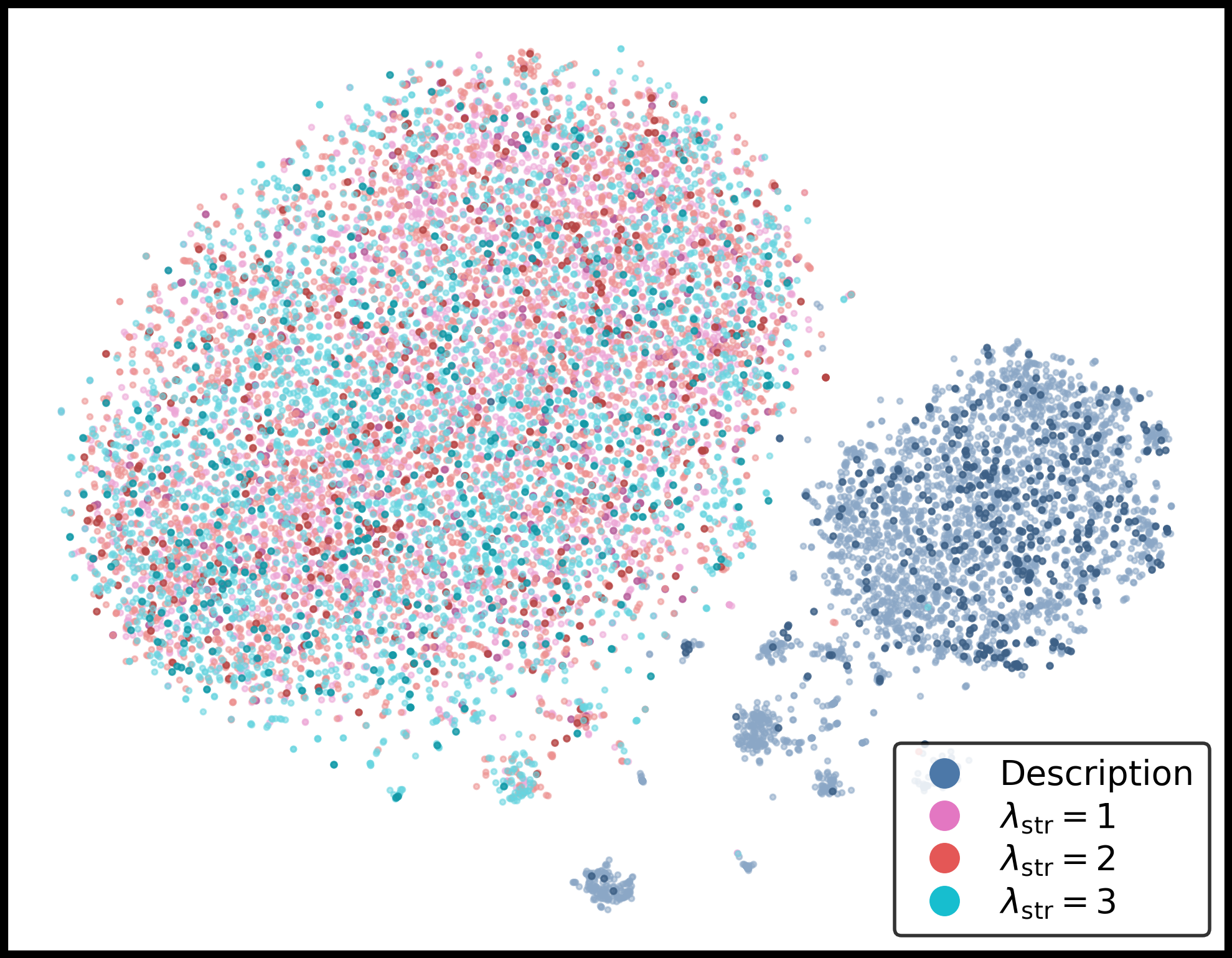}
    \caption{Varying $R_{\mathrm{str}}$ weight.}
    \label{fig:sub_add_2}
\end{subfigure}
\hspace{0.05\linewidth}% 
\begin{subfigure}{0.25\linewidth}
    \centering
    \includegraphics[width=\linewidth]{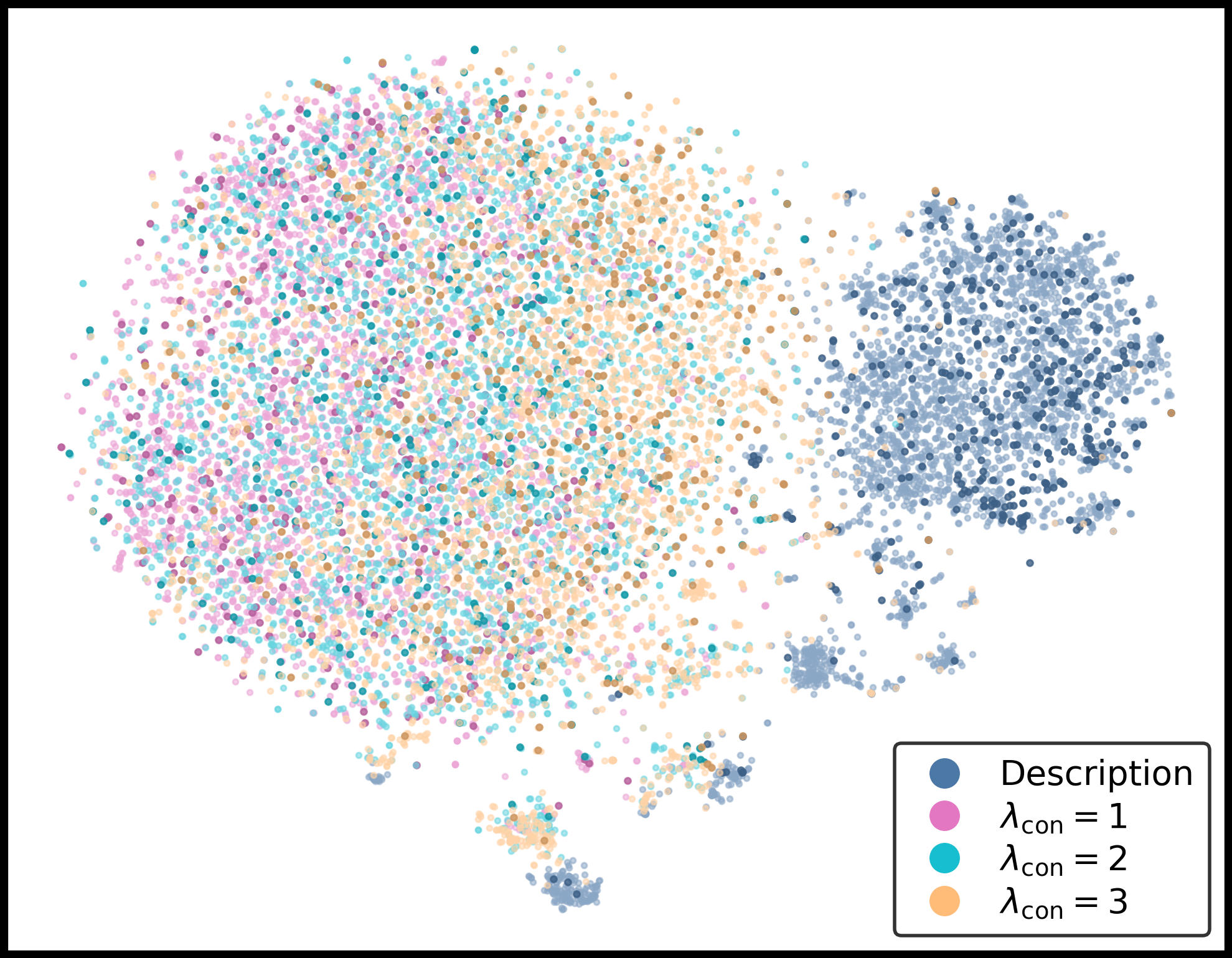}
    \caption{Varying $R_{\mathrm{con}}$ weight.}
    \label{fig:sub_add_3}
\end{subfigure}
\caption{T-SNE visualizations of embeddings under different reward weight configurations.}
\label{fig:three_subplots}
\end{figure*}

\subsection{Reward Weight Configuration}
\nosection{Individual Dissimilarity Reward} 
Fig.~\ref{fig:sub_add_1} shows that increasing the weight of the individual dissimilarity reward noticeably shifts the embedding distribution. 
As shown in Table~\ref{tab:weight_psa} and Fig.~\ref{fig:three_subplots}, larger dissimilarity weights substantially reduce CosSim, indicating stronger semantic displacement and improved privacy protection. 
The accompanying increase in D-1 and D-2 further suggests that this reward promotes more diverse variants. 
However, excessive emphasis on dissimilarity slightly increases NLL and reduces Spearman correlation, reflecting a mild loss in fluency and local structural preservation. 
Overall, this reward mainly controls privacy strength by pushing variants away from their originals while keeping downstream performance relatively stable.

\nosection{Structural Consistency Reward} 
Fig.~\ref{fig:sub_add_2} shows that the structural consistency reward mainly affects local sample positions rather than the global embedding layout. 
As reported in Table~\ref{tab:weight_psa} and Fig.~\ref{fig:three_subplots}, increasing its weight improves Spearman correlation, confirming that this reward helps preserve local semantic topology. 
Meanwhile, D-1 and D-2 increase slightly, and NLL rises only moderately, suggesting that the model can strengthen relational consistency without severely sacrificing diversity or fluency. 
Although CosSim increases slightly because the relative contribution of dissimilarity is reduced, the overall privacy--utility balance remains stable. 
Thus, this reward primarily controls relational fidelity by maintaining neighborhood structure.

\nosection{Semantic Constraint Reward} 
Fig.~\ref{fig:sub_add_3} shows that the semantic constraint reward affects both global layout and local sample positions. 
As shown in Table~\ref{tab:weight_psa} and Fig.~\ref{fig:three_subplots}, increasing its weight encourages generated variants to stay closer to the category-specific semantic manifold, improving semantic validity and preventing degenerate outputs. 
However, overly strong semantic constraints make the variants closer to the original words, as reflected by higher CosSim and reduced diversity. 
This weakens privacy protection because the individual dissimilarity objective becomes relatively less influential. 
Therefore, the semantic constraint reward is most effective when used as a regularizer that preserves validity without overly suppressing semantic shift.

\subsection{Further Observation}
Across the above analyses, we further observe that once the local semantic skeleton is sufficiently preserved, enhancing fine-grained semantic fidelity brings limited gains to recommendation performance. 
This supports our relational fidelity principle: effective recommendation does not require literal semantic preservation, but rather a stable relative structure among item attributes. 
The remaining utility is likely supported by recommender model and complementary collaborative signals.
\end{document}